# Designing Bimetallic Nanoparticle Catalysts via Tailored Surface Segregation

Yaxin Tang[1,2,†], Mingao Hou[1,†], Qian He[2,*], and Guangfu Luo[1,3,4*]

[1]Department of Materials Science and Engineering, Southern University of Science and Technology, Shenzhen 518055, China

[2]Department of Material Science and Engineering, College of Design and Engineering, National University of Singapore, 9 Engineering Drive 1, EA #03-09, 117575, Singapore

[3]Guangdong Provincial Key Laboratory of Computational Science and Material Design, Southern University of Science and Technology, Shenzhen 518055, China

[4]Institute of Innovative Materials, Southern University of Science and Technology, Shenzhen 518055, China

[†]These authors contributed equally to this work

*Corresponding author E-mail: (Q.H.) mseheq@nus.edu.sg; (G.L.) luogf@sustech.edu.cn

## ABSTRACT

Bimetallic nanoparticles serve as a vital class of catalysts with tunable properties suitable for diverse catalytic reactions, yet a comprehensive understanding of their structural evolution under operational conditions as well as their optimal design principles remains elusive. In this study, we unveil a prevalent surface segregation phenomenon in approximately 100 platinum-group-element-based bimetallic nanoparticles through molecular dynamics simulations and derive a thermodynamic descriptor to predict this behavior. Building on the generality and predictability of surface segregation, we propose leveraging this phenomenon to intentionally enrich the nanoparticle surface with noble-metal atoms, thereby significantly reducing their usage while maintaining high catalytic activity and stability. To validate this strategy, we investigate dozens of platinum-based bimetallic nanoparticles for propane dehydrogenation catalysis using first-principles calculations. Through a systematic examination of the catalytic sites on nanoparticle surfaces, we eventually identify several candidates featuring with stable Pt-enriched surface and superior catalytic activity, confirming the feasibility of this approach.

Bimetallic nanoparticle catalysts offer unique advantages over their monometallic counterparts, such as high catalytic site utilization, tunable electronic properties, and cost-effectiveness.[1–4] These attributes have led to their wide applications in various reactions, such as dehydrogenation,[5,6] hydrogenation,[7–9] hydrogen evolution,[10,11] oxygen reduction,[12–14] and carbon dioxide reduction.[15,16] Despite these advantages, the actual structures of bimetallic nanoparticles under operating conditions remains elusive.[17] Previous studies occasionally suggest that certain bimetallic nanoparticles[18–21] or dilute impurities in transition metal hosts[22,23] exhibit uneven composition distribution, where one component preferentially accumulates on the surface. For instance, Pt-Ni nanoparticles were found to exhibit a platinum-rich frame,[24] while Pt-Sn nanoparticles possess a tin-rich surface.[25,26] The driving force behind this structural evolution were previously attributed to factors, such as atomic radius and electronegativity[27,28]. Nevertheless, these studies are limited by either the narrow range of bimetallic alloys investigated or by treating one element as a dilute impurity rather than an alloy component. As a result, the generality of this phenomenon and the fundamental mechanisms governing it remain poorly understood, which hinders the development of effective countermeasures and optimal design principles for these nanoparticles. Meanwhile, the theoretical design of bimetallic catalysts frequently relies on simulations over ideal crystalline surfaces,[29–32] overlooking potential structural deviations under operational conditions. Therefore, a systematic study of the structural evolution, exploration of underlying mechanisms, and proposal of rational design principles are crucial for advancing bimetallic nanoparticle catalysts.

In this study, we investigate the structural evolution of 102 platinum-group-element-based bimetallic nanoparticles using first principles-based machine-learning molecular dynamics simulations. Our findings indicate that surface segregation is a universal phenomenon present in almost all bimetallic nanoparticles we examined. We further derive a thermodynamic descriptor of this phenomenon using surface energies of the two pure components and formation energy of the bulk alloy. Building upon this insight, we propose designing bimetallic nanoparticle catalysts by intentionally exploiting surface segregation. To validate this strategy, we computationally explore 6 platinum-based bimetallic nanoparticles for propane dehydrogenation (PDH) catalysis and successfully identify several candidates with outstanding catalytic performances.

To explore potential structural evolutions in bimetallic nanoparticles, we employ molecular dynamics (MD) simulations with on-the-fly machine learning force fields[33] for 102 platinum-group-element (Pt, Pd, Ir, Ru, Rh, and Os) based bimetallic nanoparticles (see computational details in the Supporting information). The nanoparticles, with volumes ranging from approximately 0.34 to 1.85 $nm^3$, are constructed in a near-spherical shape from their bulk materials while maintaining the bulk stoichiometries. These alloys are chosen due to the extensive use of platinum-group elements in catalysis and the experimental availibility of their bulk forms. Our MD simulations are conducted at 873 or 1773 K with a time step of 1 fs for 50 picoseconds. The MD trajectories demonstrate dynamic bond breaking and reformation throughout the simulations, with the free energies showing an overall decreasing trend (Figure S3). Upon reaching thermal equilibrium, most nanoparticles exhibit notable surface segregation (Figures S4-S11). As exemplified in Figure 1A, the $PtGa_3$ nanoparticle displays a surface accumulation of Ga, while Pt atoms predominantly reside within the nanoparticle. Given that surface atoms on a nanoparticle exhibit much smaller solid angles with respect to other atoms than those embedded within, we utilize solid angle to characterize surface segregation (see details in the Supporting information). Statistical analyses of solid angles during MD simulations confirm the surface accumulation of Ga atoms, with the Ga peak significantly preceding the Pt peak (Figure 1B). By sharp contrast, the $PtTi_3$ nanoparticle exhibits the surface accumulation of Pt atoms (Figure 1C), despite its much lower atomic ratio. This phenomenon is also supported by the statistical analyses of solid angles (Figure 1D). Our analyses at different temperatures reveal that these surface segregation phenomena persist consistently, highlighting their thermodynamic nature (Figure S12). Because of the contrasting surface segregation behaviors in $PtTi_3$ and $PtGa_3$ nanoparticles, the number of exposed Pt atoms in $PtTi_3$ is approximately twice that in $PtGa_3$, with around 16 and 8 of Pt atoms on the surface, respectively.

Further analyses reveal that the surface segregation in bimetallic nanoparticles is primarily determined by the surface energies of the constitute elemental bulks and the formation enthalpy of the bimetallic bulk. Figure 1E maps the surface segregation of each bimetallic nanoparticle with respect to the lowest surface energies of the two elemental bulks forming the alloy (see a complete list in Table S1). The red dots highlight nanoparticles with an accumulation of the platinum-group elements on the surface, which is advantageous for optimizing the use of these precious elements, while the blue dots represent nanoparticles where platinum-group elements are buried within. Figure 1E indicates that the component with a lower surface energy generally tends to segregate to the nanoparticle surface. Although this may seem intuitive from an energy minimization perspective, it is somewhat surprising given that the segregated surface of a bimetallic nanoparticle differs significantly from an ideal crystal surface. We also observe that several nanoparticles, notably PdAl, PtAl, RhBe, and IrTi, deviate from this general trend, with the lower-surface-energy component buried inside. Since surface segregation represents a departure from the alloy state, the alloy's bulk stability is supposed to play a role in this phenomenon. Our further analysis reveals that by comparing the difference in crystal surface energy to the bulk formation enthalpy of an alloy, we can effectively predict the surface segregation for nearly all the nanoparticles studied, as illustrated in Figure 1F.

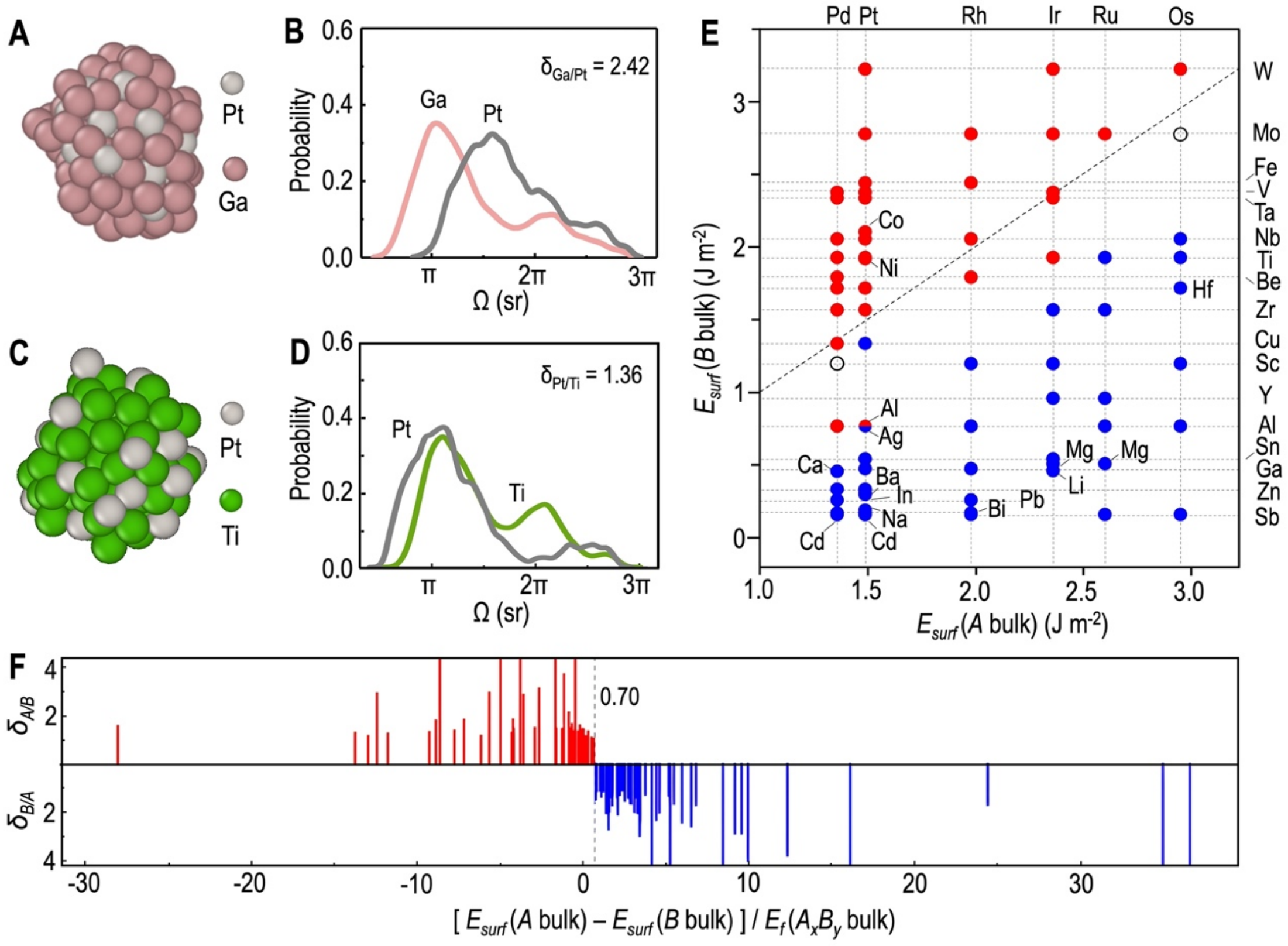

**Figure 1.** Surface segregation of platinum-group-element-based bimetallic nanoparticles under thermal equilibrium. (A) Equilibrated structure of a $PtGa_3$ nanoparticle with 112 atoms and (B) statistics of the solid angle for all atoms across 15,000 MD structures. (C) Equilibrated structure of a $PtTi_3$ nanoparticle with 96 atoms and (D) statistics of the solid angle for all atoms across 15,000 MD structures. (E) Surface segregation map of nanoparticle $A_xB_y$ with respect to the lowest surface energy of the constitute elemental bulks, namely, $E_{surf}(A$ bulk) and $E_{surf}(B$ bulk); red and blue dots represnet accumulation of platinum-group-element $A$ and promoter element $B$ on the surface, respectively, while these exhibiting no evident surface segregation, defined as a surface segregation level below 1.05, are circled. (F) Surface segregation level relative to the ratio of surface energy difference, $E_{surf}(A$ bulk) – $E_{surf}(B$ bulk), to the formation energy of $A_xB_y$ bulk, $E_f(A_xB_y$ bulk). The segregation level is defined as $\delta_{A/B} = r_{A/B}^{surf} / r_{A/B}^{bulk}$, where $r_{A/B}^{surf}$ and $r_{A/B}^{bulk}$ represent the atomic ratio of $A/B$ in the nanoparticle surface and bulk alloy, respectively.

This paramter quantifies the sotichiometry deviation in the surface layer relative to the bulk (see Supporting Information for more details).

To investigate the sensitivity of surface segregation to the stoichiometry and size of bimetallic nanoparticles, we first examine 30 platinum-based bimetallic nanoparticles with stoichiometries of $Pt_3M$, Pt$M$, and $PtM_3$ ($M$ = In, Zn, Ga, Sn, Ag, Ni, Ti, Co, Fe, and Mo). Figure 2 illustrates that all three stoichiometries exhibit the same qualitative result: the element with lower surface energy tends to accumulate on the surface. Specifically, In, Zn, Ga, Sn, and Ag segregate to the surface (Figure 2A), while Ni, Ti, Co, Fe, and Mo tend to reside inside the nanoparticles, allowing Pt to accumulate on the surface (Figure 2B). The statistics of solid angle distribution further confirms the observation (Figures S6 and S7).

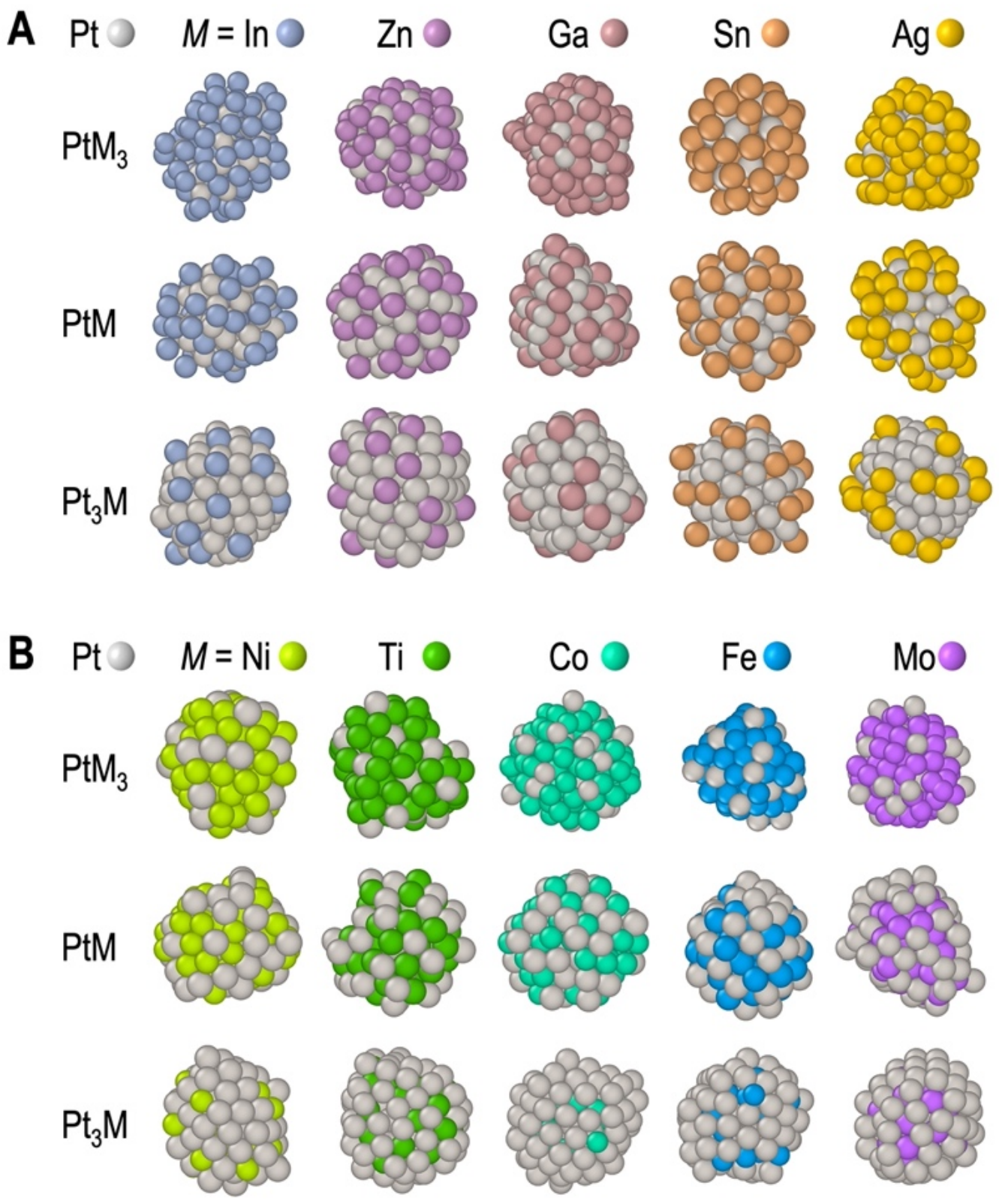

**Figure 2.** Stoichiometry dependence of surface segregation for 30 equilibrated platinum-based bimetallic nanoparticles with volumes of ~0.75 $nm^3$. Structures of nanoparticles with surface segregation of (A) promoter element or (B) platinum.

We also assess the size dependence of surface segregation by examining three volumes of ~0.34, ~0.75, and ~1.85 $nm^3$ for the $PtGa_3$, PtGa, $PtTi_3$, and PtTi nanoparticles. As depicted in Figure 3, $PtGa_3$ and PtGa nanoparticles consistently exhibit Ga segregation on the surface for all three sizes, aligning with Ga bulk's lower surface energy. Similarly, the $PtTi_3$ and PtTi nanoparticles of different sizes always show Pt accumulation on the surface. Therefore, the surface segregation is insensitive to both stoichiometry and size of bimetallic nanoparticles within our examined range. An even larger nanoparticle may not alter this trend, since the surface segregation is majorly determined by bulk properties (Figure 1E-1F), which represent the behavior of infinitely large nanoparticles.

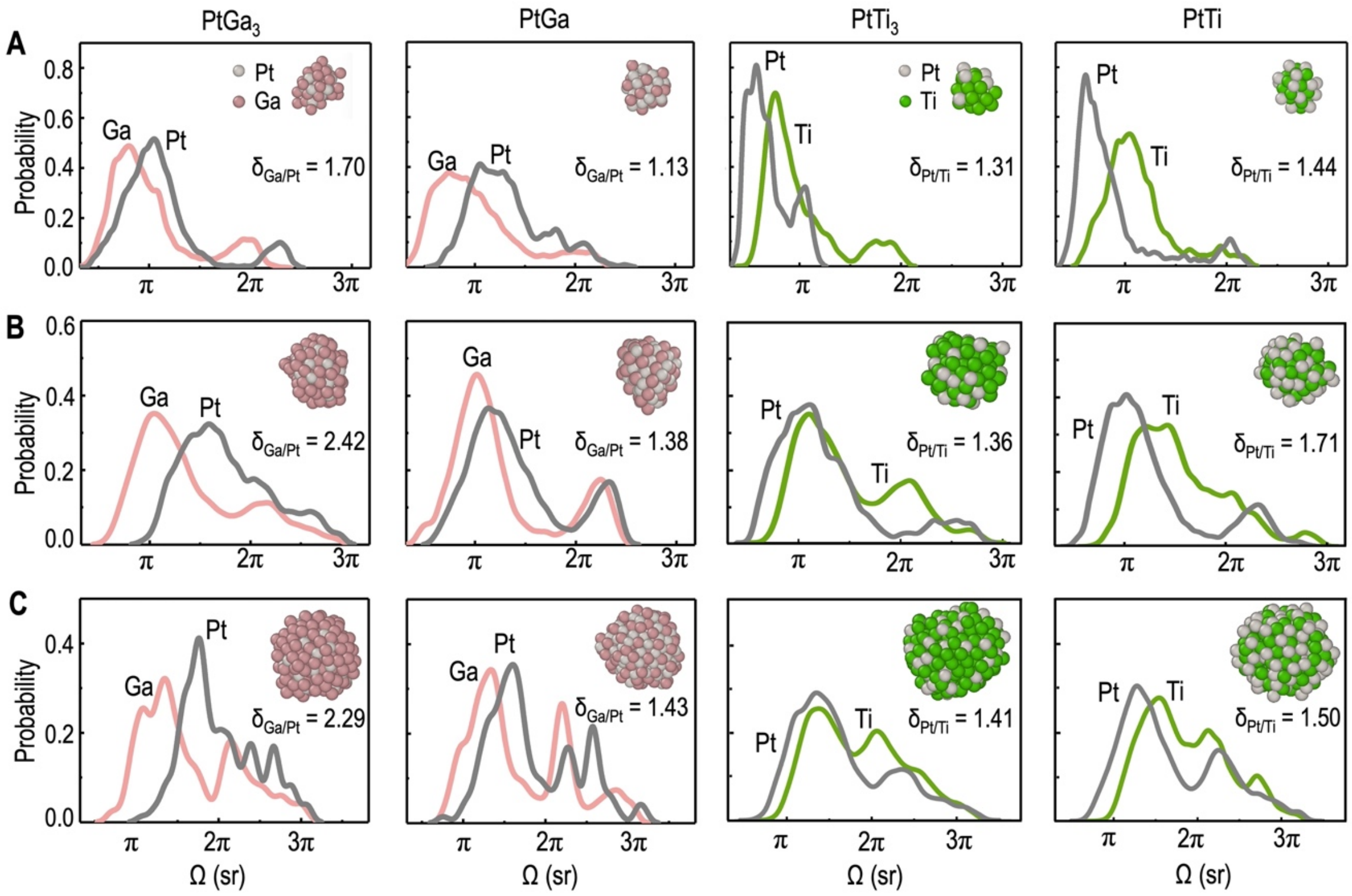

**Figure 3.** Size dependence of surface segregation for equilibrated Pt-Ga and Pt-Ti nanoparticles. Statistics of solid angle distribution for nanoparticles with volumes of (A) ~0.34, (B) ~0.75, and (C) ~1.85 $nm^3$ across 15,000 MD structures. Inset shows a snapshot of the structure under equilibrium. The the segregation level is indicated in each plot.

Because of surface segregation, the catalytic sites of nanoparticles differ significantly from those based on idealized crystal surfaces.[34] For promoter elements with both low surface energy and high volatility, the latter of which usually corresponds to low melting temperature, such as Zn in the Pt-Zn nanoparticles, surface segregation would lead to component loss and catalyst deactivation under elevated temperatures, as observed experimentally.[6]

To optimize the design of bimetallic nanoparticles by leveraging the surface segregation phenomenon, one can strategically select promoter elements characterized by relatively high surface energy and melting temperatures. This approach facilitates the segregation of catalytic elements to the nanoparticle surface while minimizing component loss. To demonstrate this strategy, we investigate the PDH catalytic activities of the $PtM_3$ ($M$ = Ni, Ti, Co, Fe, and Mo) nanoparticles, which exhibit surface accumulation of Pt, as shown in Figure 2B. Generally, the catalysis on bimetallic catalysts could involve multiple active sites, a phenomenon known as the ensemble effect.[35–37] Since Pt sites are known to play a pivotal role in the rate-determining dehydrogenation process, we first assess the catalytic activity of various Pt sites on the $PtTi_3$ surface (Figure 4A). Non-Pt sites, which may contribute to the desorption process (Figure S13), are not considered in this study. Bader charge analysis reveals that the surface Pt sites exhibit change states ranging from -1.8 to -3.3. In general, Pt sites with higher coordination numbers to Ti atoms roughly exhibit more negative charges and lower $d$-band centers (Table S2).

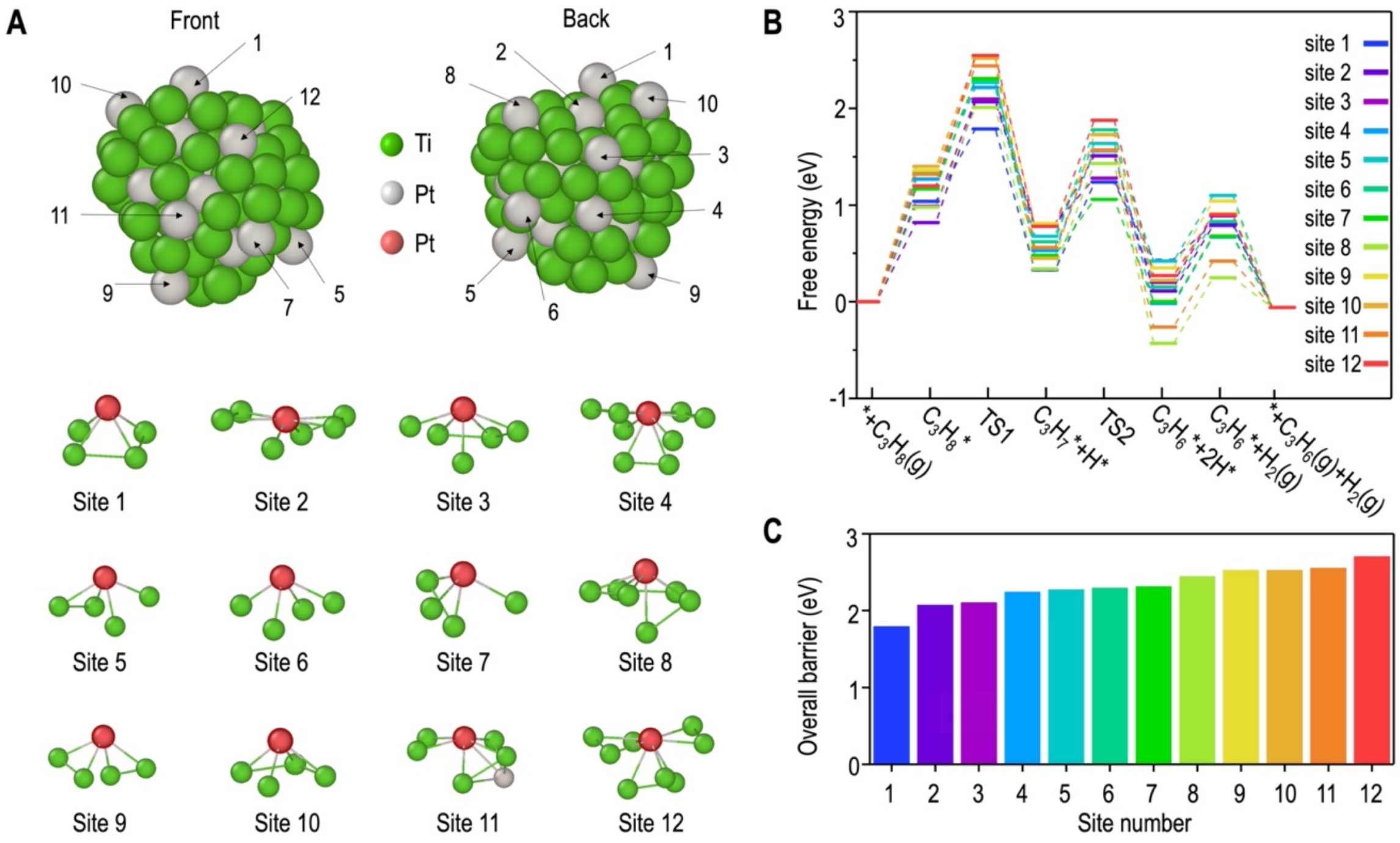

**Figure 4.** Catalytic activity of PDH reaction on an equilibrated $PtTi_3$ nanoparticle with 96 atoms. (A) Overall geometry and local structures of Pt surface sites; (B) free energy diagrams and (C) overall barriers on different surface Pt sites of a structure obtained at 873 K. Original data in panel B is provided in Table S3.

The free energy diagrams in Figure 4B indicate that the first dehydrogenation step is rate-determining for all 12 surface Pt sites, with overall energy barrier ranging from 1.79 to 2.70 eV (Figure 4C). Notably, site 1, which is geometrically elevated, exhibits the lowest energy barrier. This site demonstrates strong binding with both $C_3H_8$ and $C_3H_7$, a feature that is known to facilitate dehydrogenation. Sites 2 to 10 and 12 exhibit structures similar to site 1, with energy barriers ranging from 2.1 to 2.6 eV. Site 11 represents another type of local structure with neighboring Pt atom, which is named as “Pt ensemble site” hereafter. This site binds relatively weakly with $C_3H_8$ and $C_3H_7$, resulting in a significant barrier of 2.5 eV for the first dehydrogenation. Unlike the Pt nanoparticle (Figure 5A), which is characterized by high $C_3H_6$ desorption energy and coke formation due to strong binding strength, $C_3H_6$ desorption is exothermic for all surface Pt sites on the $PtTi_3$ nanoparticle, suggesting a potentially high selectivity for this product.[38] It is worth noting that $Pt_3Ti$ nanoparticles have previously demonstrated outstanding catalytic activity for PDH, with an overall energy barrier of 2.30 eV.[39] Our study here suggests that $PtTi_3$ may offer more efficient Pt utilization while maintaining similar catalytic activity for PDH.

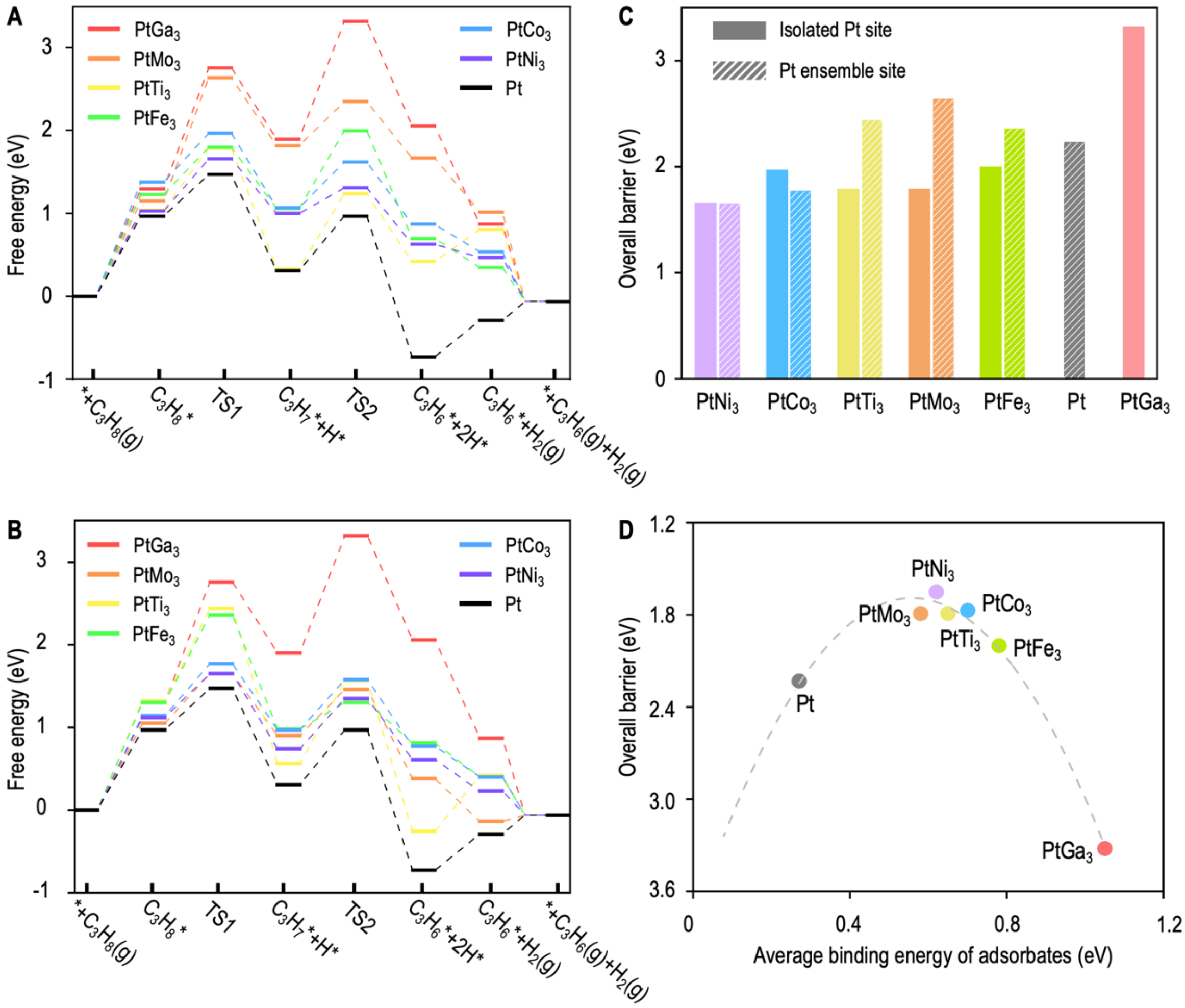

**Figure 5.** Catalytic activities of Pt$M_3$ ($M$ = Ni, Ti, Co, Fe, Mo, and Ga) nanoparticles and Pt nanoparticle for PDH under 873 K. Free energy diagrams of (A) isolated Pt sites and (B) Pt ensemble sites; and (C) comparison of overall energy barriers. (D) Correlation between overall energy barriers and the average binding energy of adsorbent $C_3H_8$, $C_3H_7$, $C_3H_6$, and H. The structure of isolated Pt sites and Pt ensemble sites are provided in Figure S14. Original data in panels A and B is provided in Table S4.

Building on our understanding of the catalytic sites on the $PtTi_3$ nanoparticle, we extend our evaluation of PDH catalysis to the other nanoparticles: $PtNi_3$, $PtCo_3$, $PtFe_3$, and $PtMo_3$. Given the local structures of surface Pt sites on the $PtTi_3$ nanoparticle, we focus on two types of surface configurations: isolated Pt sites in geometrically elevated positions and Pt ensemble sites with neighboring Pt atoms, which correspond to sites 1 and 11 in Figure 4A, respectively. For comparison, we also assess the $PtGa_3$ nanoparticle, which exhibits surface accumulation of Ga (Figures 2A and 3B), and the Pt nanoparticle. As anticipated, the Pt nanoparticle demonstrates strong binding with all intermediate states, resulting in low activation barriers for both dehydrogenation steps. However, this strong binding hinders $C_3H_6$ desorption, leading to a significant overall barrier (Figure 5A) and promoting further dehydrogenation and coke formation.[38] By contrast, the $PtGa_3$ nanoparticle binds too weakly to the intermediate states, as most Pt atoms are buried within the structure, resulting in very high dehydrogenation barriers exceeding 3 eV.

The other nanoparticles exhibit binding strength fall between those of Pt and $PtGa_3$ nanoparticles, resulting in lower overall energy barriers. For isolated Pt sites, $PtNi_3$ demonstrates the lowest barrier of 1.66 eV, followed by $PtTi_3$ (1.79 eV), $PtCo_3$ (1.97 eV), $PtFe_3$ (2.0 eV), and $PtMo_3$ (2.64 eV), as illustrated in Figure 5A and 5C. For Pt ensemble sites, $PtNi_3$ again shows the lowest barrier of 1.65 eV, followed by $PtCo_3$ (1.77 eV), $PtMo_3$ (1.79 eV), $PtFe_3$ (2.36 eV), and $PtTi_3$ (2.50 eV), as shown in Figure 5B and 5C. Overall, $PtNi_3$, $PtCo_3$, $PtTi_3$, and $PtMo_3$ nanoparticles exhibit the top four catalytic abilities for PDH. Their exceptional activities can be attributed to an optimal balance in Pt binding strength (Figure 5D), which is neither too strong, as in Pt, nor too weak, as in $PtGa_3$. This characteristic remains consistent across different temperatures, while the overall energy barriers tend to decrease with temperature (Figure S15).

Finally, we would like to briefly discuss the potential for experimental verification. The proposed catalysts could be synthesized using established methods for Pt-Co[18] and Pt-Sn[26] nanoparticles, such as co-precipitation, sol-gel processing, and colloidal synthesis. Key considerations for successful synthesis may include controlling stoichiometry, managing synthesis temperature, and minimizing oxidation during this process. Since surface segregation is thermodynamically driven, this phenomenon is expected to remain robust under experimental conditions, as evidenced by the observed surface segregation in Pt-Co and Pt-Sn nanoparticles.[18,26] Nevertheless, if the thermodynamic driving force is weak, which corresponds to the region near the boundary in Figure 1F, the adsorption of specific reaction species may also influence the surface segregation.

In summary, this study unveils a prevalent surface segregation phenomenon in platinum-group-element-based bimetallic nanoparticles. This phenomenon can largely be predicted by the crystal surface energies of the two components and the formation enthalpy of bimetallic alloy, with the lower-surface-energy component typically accumulating on the surface. Moreover, it appears to be insensitive to both stoichiometry and nanoparticle size within the examined ranges. To enhance the utilization of catalytic noble-metal elements and minimize component loss, we propose selecting a promoter with high surface energy and a relatively high melting temperature. We validate this strategy by applying it to a series of platinum-based bimetallic nanoparticles for PDH catalysis and identify $PtNi_3$, $PtCo_3$, $PtTi_3$, and $PtMo_3$ as the top four high-performance catalysts, showcasing high Pt utilization, potentially high stability, and low overall energy barrier. Given the thermodynamical nature of surface segregation, the insights from this study may offer a potential framework for the optimal design of bimetallic nanoparticles more broadly.

**ASSOCIATED CONTENT**

**Supporting Information**

The Supporting Information is available free of charge on the ACS Publications website; Computational details, details of solid angle distribution, energy and temperature evolution during MD simulations, solid angle distributions for all examined nanoparticles, solid angle distributions at different MD simulation temperatures, comparison of $C_3H_6$ desorption energies at Pt and non-Pt sites, detailed parameters used to predict surface segregation, charge states and *d*-band centers for Pt sites in $PtTi_3$ nanoparticle, values of free energy diagrams, structures of examined catalytic sites, average adsorption energy versus overall energy barriers at different temperatures (PDF)

Structures before and after MD simulations (zip)

**Notes**

The authors declare no competing interest.

## ACKNOWLEDGEMENT

This work was supported by the fund of the National Foundation of Natural Science, China (No. 52273226), the Guangdong Provincial Key Laboratory of Computational Science and Material Design (No. 2019B030301001), the Shenzhen Science and Technology Innovation Commission (No. JCYJ20200109141412308), and the High level of special funds of Southern University of Science and Technology (No. G03050K002). Q.H. was supported by the National Research Foundation Singapore (NRF-NRFF11-2019-0002). All the calculations were carried out on the Taiyi cluster supported by the Center for Computational Science and Engineering of Southern University of Science and Technology.

**Graphical Abstract**

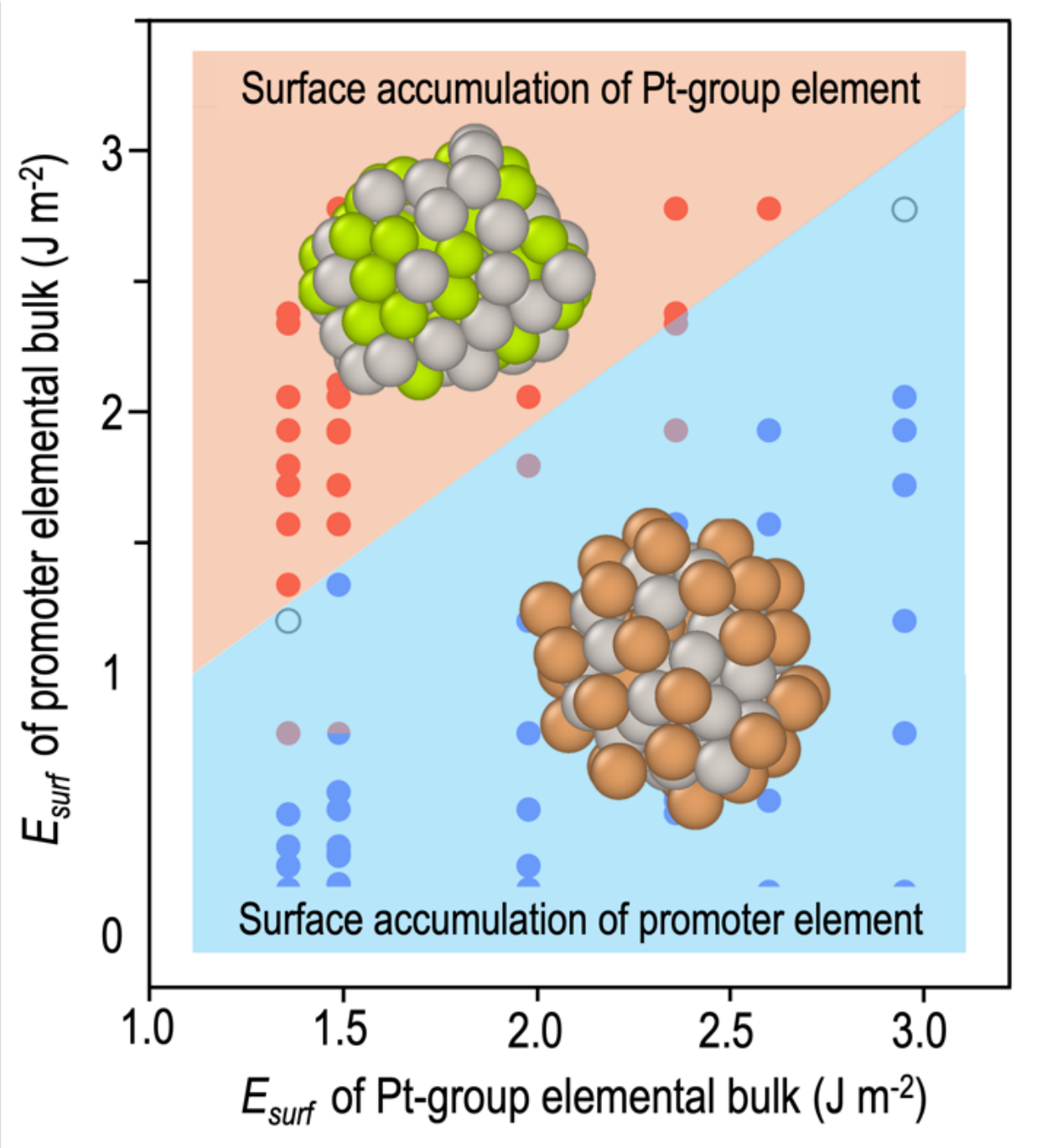

**Supporting information for "Designing Bimetallic Nanoparticle Catalysts via Tailored Surface Segregation"**

Yaxin Tang[1,2,†], Mingao Hou[1,†], Qian He[2,*], and Guangfu Luo[1,3,4,*]

[1]Department of Materials Science and Engineering, Southern University of Science and Technology, Shenzhen, Guangdong 518055, China

[2]Department of Material Science and Engineering, College of Design and Engineering, National University of Singapore, 9 Engineering Drive 1, EA #03-09, 117575, Singapore

[3]Guangdong Provincial Key Laboratory of Computational Science and Material Design, Southern University of Science and Technology, Shenzhen 518055, China

[4]Institute of Innovative Materials, Southern University of Science and Technology, Shenzhen 518055, China

[†]These authors contributed equally to this work

*E-mail: (Q.H.) mseheq@nus.edu.sg; (G.L.) luogf@sustech.edu.cn

## I. Computational details on first-principles-based calculations

All density functional theory (DFT) calculations in this work were conducted using the Vienna Ab-initio Simulation Package with the Perdew-Berke-Ernzerhof exchange-correlation functional.[1–3] Van der Waals corrections are accounted for using the DFT-D3 method.[4] The plane-wave energy cutoffs vary between 230 and 450 eV depending on the system. Pseudopotentials are described using the projected augmented wave method. [5] The 102 initial nanoparticle structures are constructed from the corresponding bulk alloys obtained from the Materials Project database[6], featuring a near-spherical shape while maintaining the bulk stoichiometries. All the examined alloys are experimentally available in their bulk forms. Supercell dimensions of 25 × 25 × 25 Å$^3$ are utilized to ensure a minimum distance of 10 Å between adjacent atoms in neighboring images. A single gamma point is employed for Brillouin zone sampling. The energy and force tolerances are set to $10^{-5}$ eV and 0.01 eV Å$^{-1}$, respectively.

Molecular dynamics simulations with on-the-fly machine-learning force fields[7–9] are carried out under the canonical ensemble at 873 K for most systems, while an increased temperature of 1773 K is used for systems containing Ta, W, Re, and Os, which are known for extremely high melting temperatures in elemental bulks. The simulation time is 50 ps, with a timestep of 1 fs. Structural analyses are based on the trajectories from the last 15 ps. The energy and temperature evolution during MD simulations are monitored to confirm the achievement of thermal equilibrium. Transition states during the PDH reaction are calculated using the climbing-image nudged elastic band method,[10] with 3–5 initial image structures generated by the adaptive-semirigid-body-approximation method.[11]

In the free energy diagrams, Gibbs free energy corrections are included for gaseous and adsorbate species using Eqn. S1:

$$\Delta G = \Delta H - T\Delta S\,, \tag{S1}$$

where $\Delta H$ and $\Delta S$ are enthalpy and entropy corrections, respectively, contributed by vibration for all species and additional translational contributions for gaseous species. The temperature $T$ is set to 873 K, while the partial pressures of $H_2$, $C_3H_8$, and $C_3H_6$ are 0.167, 0.67, and 0.167 atm, respectively, according to the typical conditions of PDH reactions.[12]

The overall energy barrier ($E_b$) for a reaction path is defined as the free energy difference between the highest and lowest states in a free energy diagram, as described by Eqn. S2.

$$E_b \equiv \max\{G_i\} - \min\{G_i\} \tag{S2}$$

Here, $\{G_i\}$ represents the free energies of all states in a free energy diagram, including the transition states. This parameter accounts for both the activation barrier and the occupancy of the relevant state, and it is closely related to the reaction rate, as demonstrated in the following deduction. Assuming that state $j$ corresponds to the highest state in a free energy diagram, the activation barrier from state $j$-1 to state $j$ is expressed by Eqn. S3.

$$E_a = G_j - G_{j-1} = \max\{G_i\} - G_{j-1} \tag{S3}$$

On the other hand, the occupancy of state $j$-1 is given by Eqn. S4 from the perspective of microkinetic theory.[13,14]

$$n_{j-1} \propto e^{-\frac{G_{j-1}-\min\{G_i\}}{k_B T}} \tag{S4}$$

Consequently, the overall reaction rate $r$ can be expressed as Eqn. S5.

$$r = n_{j-1}\, e^{-\frac{E_a}{k_B T}} \propto e^{-\frac{\max\{G_i\}-\min\{G_i\}}{k_B T}} \tag{S5}$$

This demonstrates that the reaction rate is governed by $\max\{G_i\} - \min\{G_i\}$, which is defined as the overall energy barrier in this study. It is evident that a smaller overall energy barrier corresponds to higher catalytic ability.

To evaluate the binding strength of a specific site, we calculate the average binding energies of $C_3H_8$, $C_3H_7$, $C_3H_6$, and H at the same site using the definitions in Eqn. S6.

$$\langle E_{binding}\rangle = [G_{ad}(C_3H_8) + G_{ad}(C_3H_7) + G_{ad}(C_3H_6) + G_{ad}(H)]/4 \tag{S6-1}$$

$$G_{ad}(C_3H_8) = G(C_3H_8^*) - G(C_3H_8) - G(*) \tag{S6-2}$$

$$G_{ad}(C_3H_7) = G(C_3H_7^*) + 0.5\, G(H_2) - G(C_3H_8) - G(*) \tag{S6-3}$$

$$G_{ad}(C_3H_6) = G(C_3H_6^*) - G(C_3H_6) - G(*) \quad \text{(S6-4)}$$

$$G_{ad}(H) = G(H^*) - 0.5\, G(H_2) - G(*) \quad \text{(S6-5)}$$

Here, the free energies are calculated under the same conditions as those for free energy diagrams.

## II. Calculation of solid angle distribution in a nanoparticle

To quantitatively characterize surface segregation in bimetallic nanoparticles, we employ the solid angle of each atom relative to all the other atoms to determine its position within the nanoparticle. Atoms located inside a nanoparticle exhibit a solid angle close to 4π, whereas surface atoms have small solid angles. This parameter represents the way humans typically differentiate between the "inside" and "outside". The procedure of solid-angle-distribution analysis consists of five steps.

1. Select MD trajectories under thermal equilibrium and extract the atomic coordinates. The last 15,000 structures are used in this work.
2. For each atom *i* in structure *n*, calculate vectors between atom *i* and all other atoms (Figure S1A); project these vectors onto the surface of a unit sphere centered at atom *i* to obtain a series of points *p*.
3. Group the points *p* into semi-continuous regions and calculate their total solid angle *Ω* relative to the sphere center.
4. Apply this method to all atoms in the selected trajectories to generate a normalized distribution probability of solid angle for each element, *P*(*Ω*), as shown in Figure S1B. Further analyses indicate that each peak in the solid angle distribution corresponds to a radial accumulation of specific atoms (Figure S1C).
5. Calculate "surface segregation level" to reflect the stoichiometry deviation in a surface layer relative to the nanoparticle according to Eqn. S4:

$$\delta_{A/B} = \frac{\int_0^{\Omega_0} P_A(\Omega)d\Omega}{\int_0^{\Omega_0} P_B(\Omega)d\Omega}\ , \quad \text{(S4)}$$

where $P_A(\Omega)$ and $P_B(\Omega)$ are the solid angle distributions, respectively, for elements *A* and *B* in a nanoparticle. The surface layer is defined by a critical solid angle $\Omega_0$, below which the number of atoms equals that in a 1.8 Å-thick shell of an equivalent sphere, as defined by Eqn. S5:

$$\frac{N_A * \int_0^{\Omega_0} P_A(\Omega)d\Omega + N_B * \int_0^{\Omega_0} P_B(\Omega)d\Omega}{N_A + N_B} = \frac{R^3 - (R-1.8\text{Å})^3}{R^3}\ , \quad \text{(S5)}$$

where $N_A$ and $N_B$ are the total number of atoms for elements *A* and *B* in the nanoparticle; *R* is the radius of an equivalent sphere that has the same number of atoms as the nanoparticle. If $1.05 \le \delta_{A/B}$, element *A* is considered to exhibit noticeable surface accumulation; A value of $1 \le \delta_{A/B} < 1.05$ suggests no significant surface segregation, while $\delta_{A/B} < 1$ indicates surface accumulation of element *B*, in which case $\delta_{B/A}$ is used instead for easy comparison.

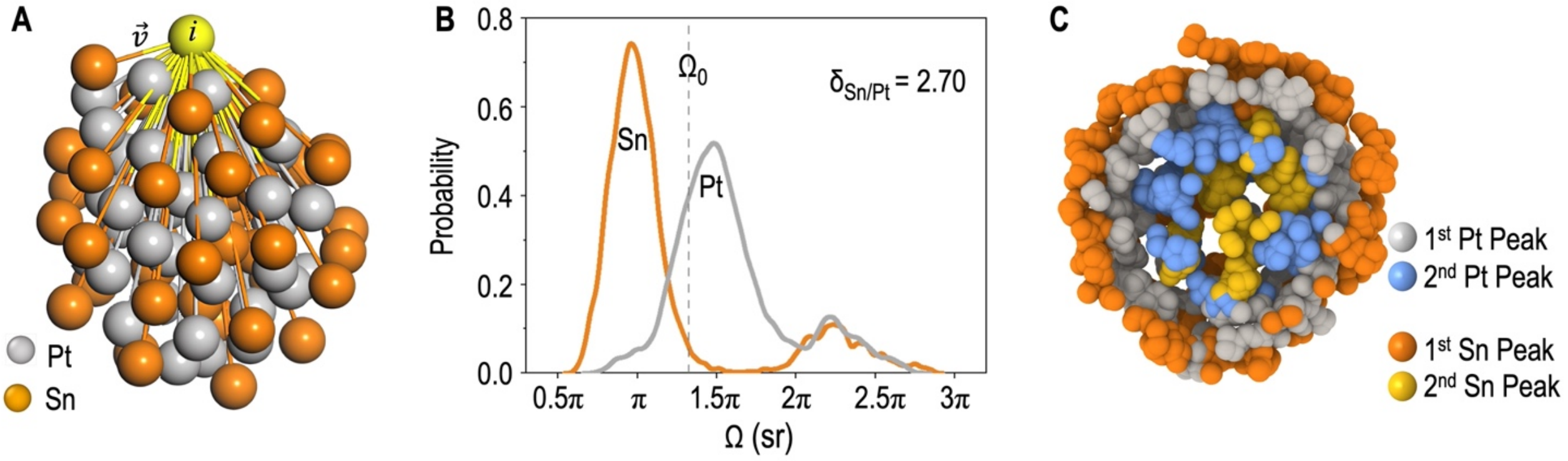

**Figure S1**. (A) Vectors representing the relative positions of a chosen atom with respect to all other atoms. (B) Solid angle distribution for a 15,000-frame trajectory of a $Pt_{41}Sn_{41}$ nanoparticle, as well as the critical solid angle $\Omega_0$ and surface segregation level $\delta_{Sn/Pt}$. (C) Superposition of atoms from the MD trajectories analyzed in panel (B), with a cross-sectional view. A correspondence is evident between the peaks in panel (B) and the radial accumulation of atoms shown in panel (C).

We also compare the structural analyses using solid angle distribution and the coordination number, the latter being another potential parameter for identifying surface atoms. As shown in Figure S2, both methods clearly reveal a surface segregation, validating the reliability of our approach. A key distinction is that a solid angle accounts for the relative positions of all atoms in the nanoparticle, whereas a coordination number is determined by local atomic configuration, which can sometimes misinterpret the relative position. For instance, an atom inside a hollow nanoparticle can be incorrectly identified as a surface atom based on coordination number. When analyzing using solid angle, the atom exhibits a solid angle close to 4π and thus is correctly classified as an inner atom.

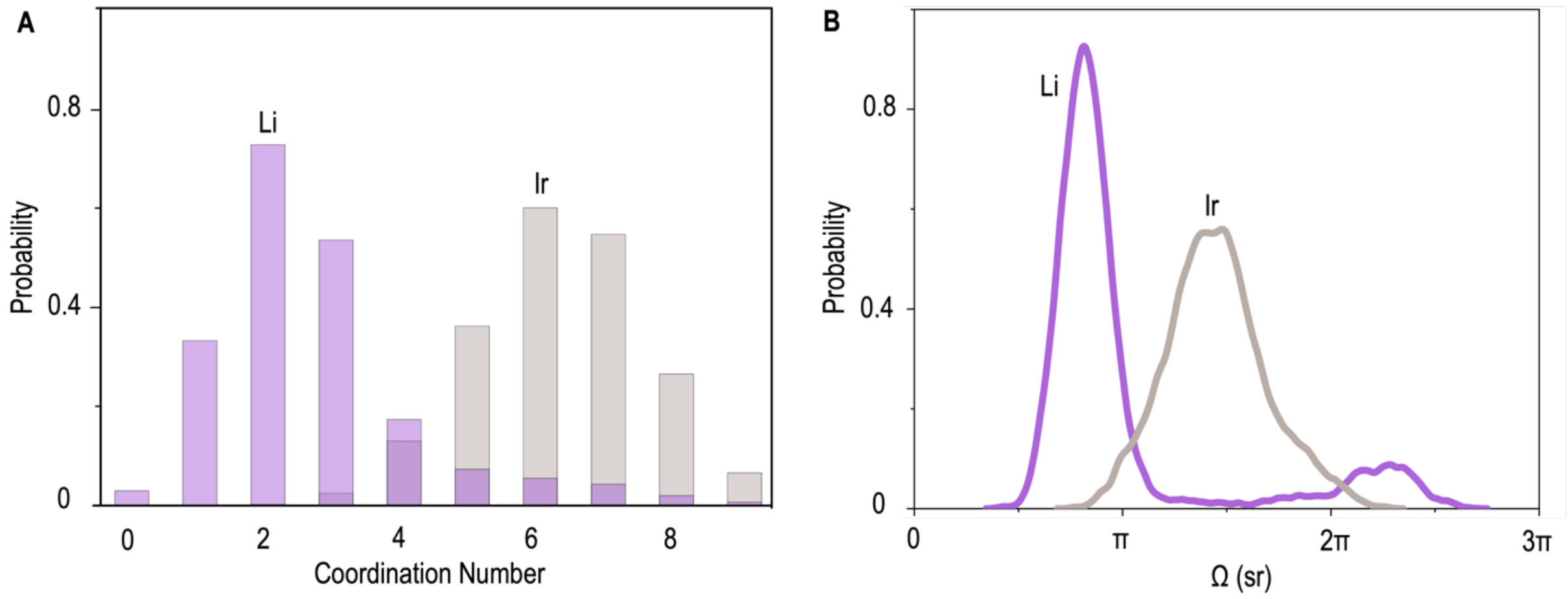

**Figure S2.** Structural analyses of the IrLi nanoparticle using (A) coordination number based on bond length analysis and (B) solid angle distribution proposed in this work. The Ir-Ir, Ir-Li, and Li-Li bonds are defined as atom pairs with distances within 2.42 ± 0.3, 2.52 ± 0.3, and 2.62 ± 0.3 Å, respectively.

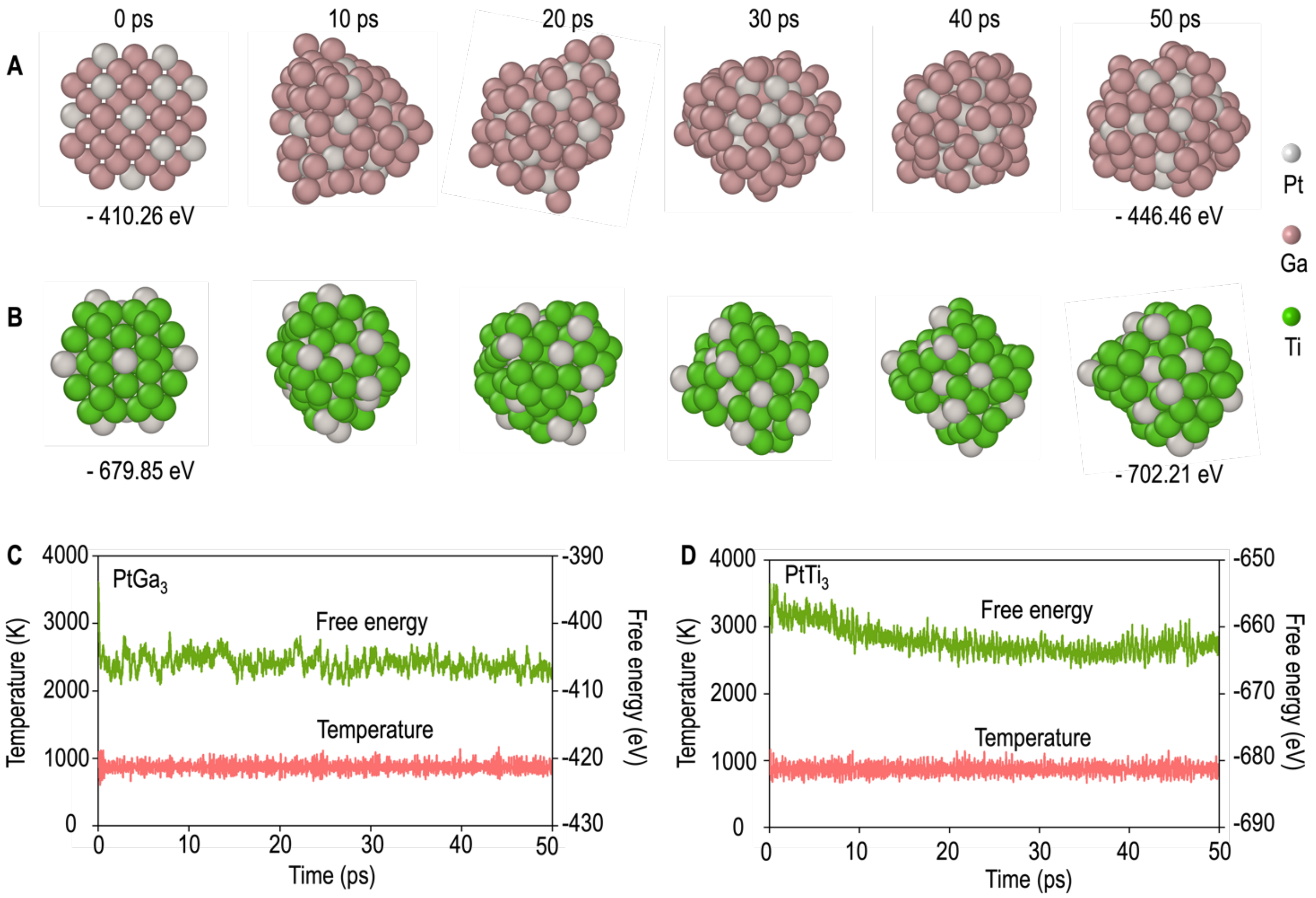

**Figure S3.** Structural snapshots taken at 0, 10, 20, 30, 40, and 50 ps during MD simulations for the (A) $PtGa_3$ and (B) $PtTi_3$ nanoparticles at 873 K. The free energies of the initial and final states are labeled at the bottom. Evolution of temperature and free energy during MD simulations for the (C) $PtGa_3$ and (D) $PtTi_3$ nanoparticles at 873 K.

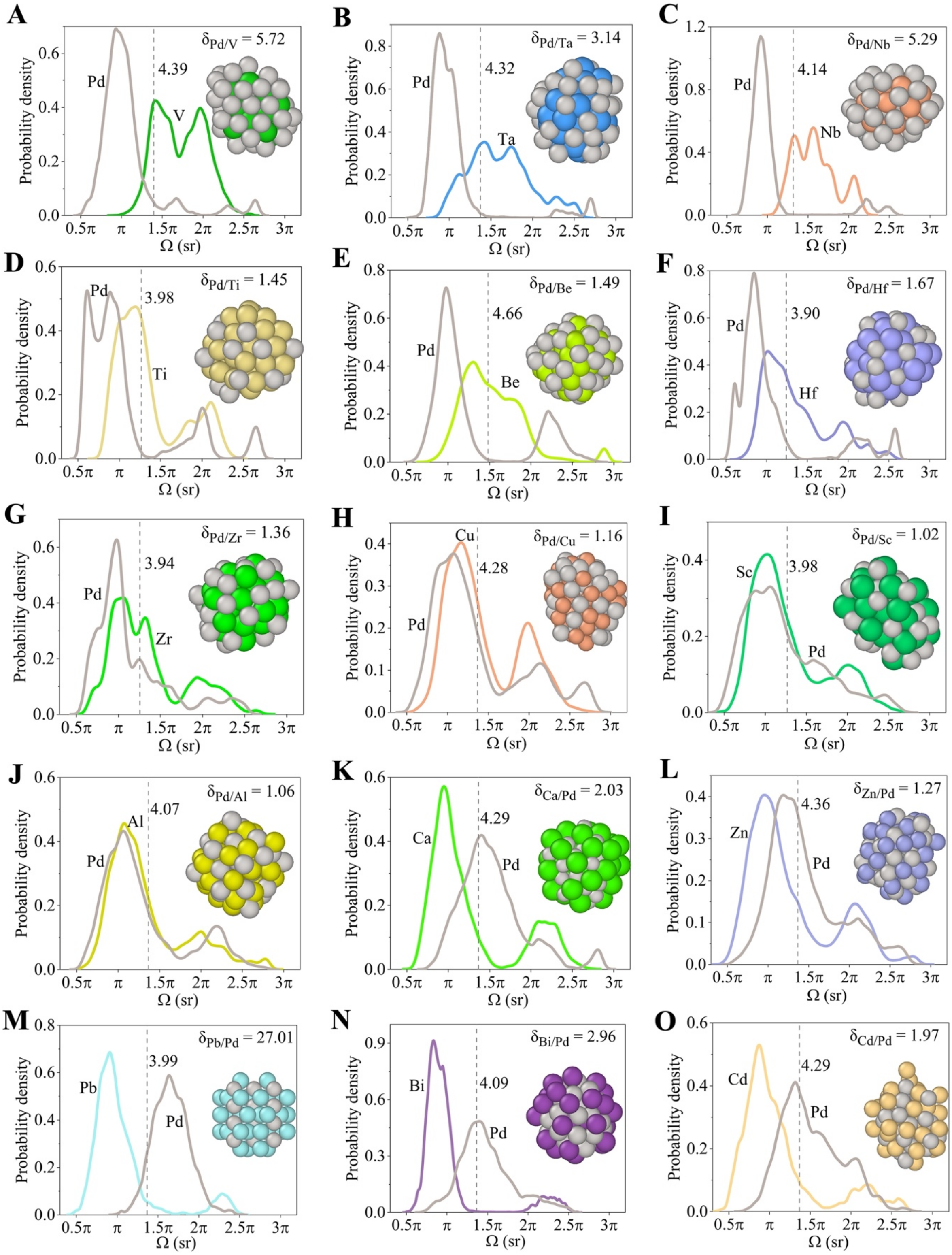

**Figure S4.** Structures at the end of MD simulations and solid angle distributions for 15 Pd-based bimetallic nanoparticles: (A) $Pd_{44}V_{22}$, (B) $Pd_{36}Ta_{36}$, (C) $Pd_{36}Nb_{18}$, (D) $Pd_{21}Ti_{42}$, (E) $Pd_{44}Be_{44}$, (F) $Pd_{21}Hf_{42}$, (G) $Pd_{30}Zr_{30}$, (H) $Pd_{44}Cu_{44}$, (I) $Pd_{32}Sc_{32}$, (J) $Pd_{37}Al_{37}$, (K) $Pd_{44}Ca_{44}$, (L) $Pd_{40}Zn_{40}$, (M) $Pd_{20}Pb_{40}$, (N) $Pd_{30}Bi_{30}$, and (O) $Pd_{36}Cd_{36}$.

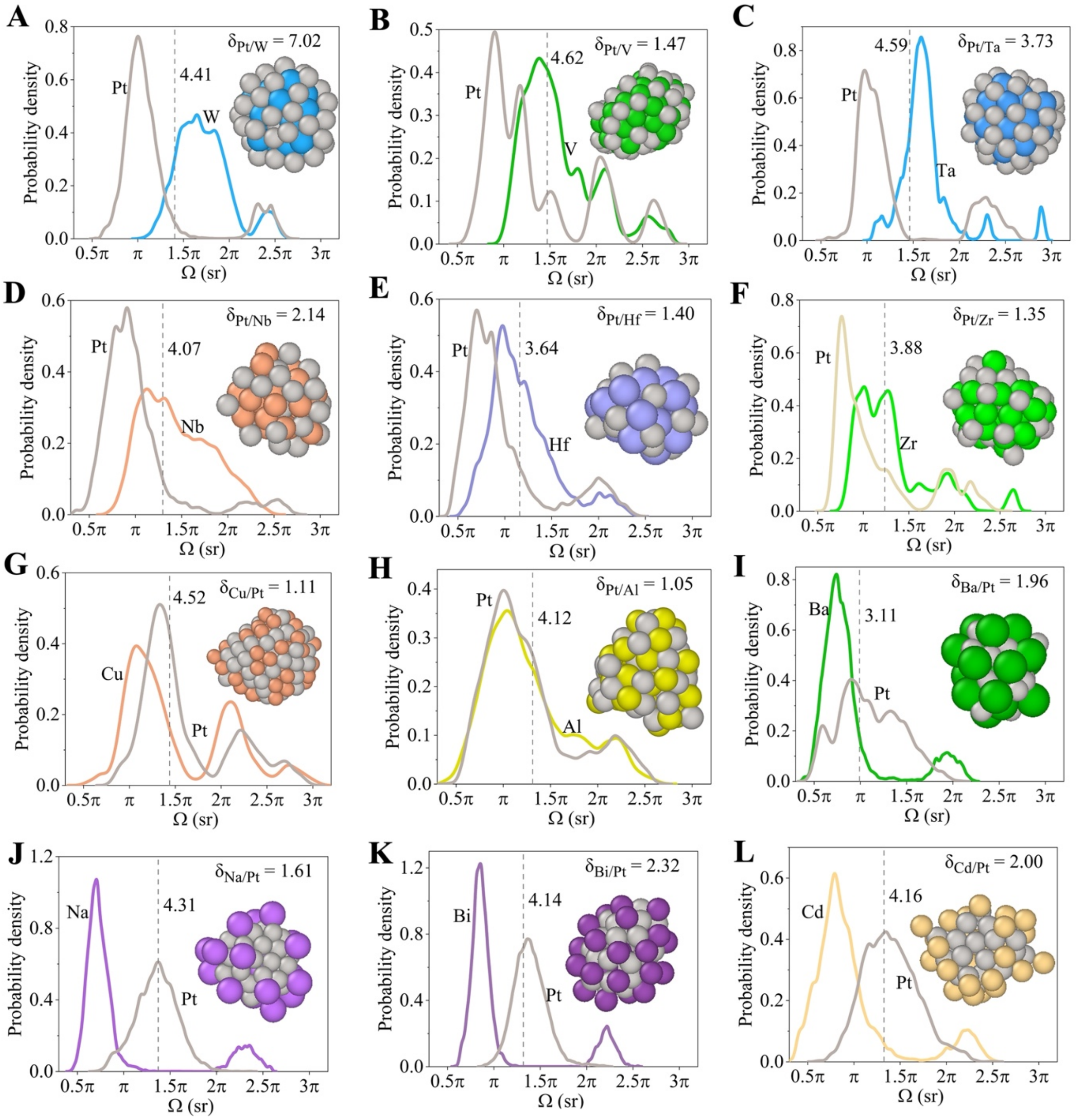

**Figure S5.** Structures at the end of MD simulations and solid angle distributions for 12 Pt-based bimetallic nanoparticles: (A) $Pt_{52}W_{26}$, (B) $Pt_{56}V_{56}$, (C) $Pt_{60}Ta_{30}$, (D) $Pt_{30}Nb_{30}$, (E) $Pt_{22}Hf_{22}$, (F) $Pt_{30}Zr_{30}$, (G) $Pt_{67}Cu_{67}$, (H) $Pt_{35}Al_{35}$, (I) $Pt_{18}Ba_{18}$, (J) $Pt_{40}Na_{20}$, (K) $Pt_{30}Bi_{30}$, and (L) $Pt_{28}Cd_{28}$.

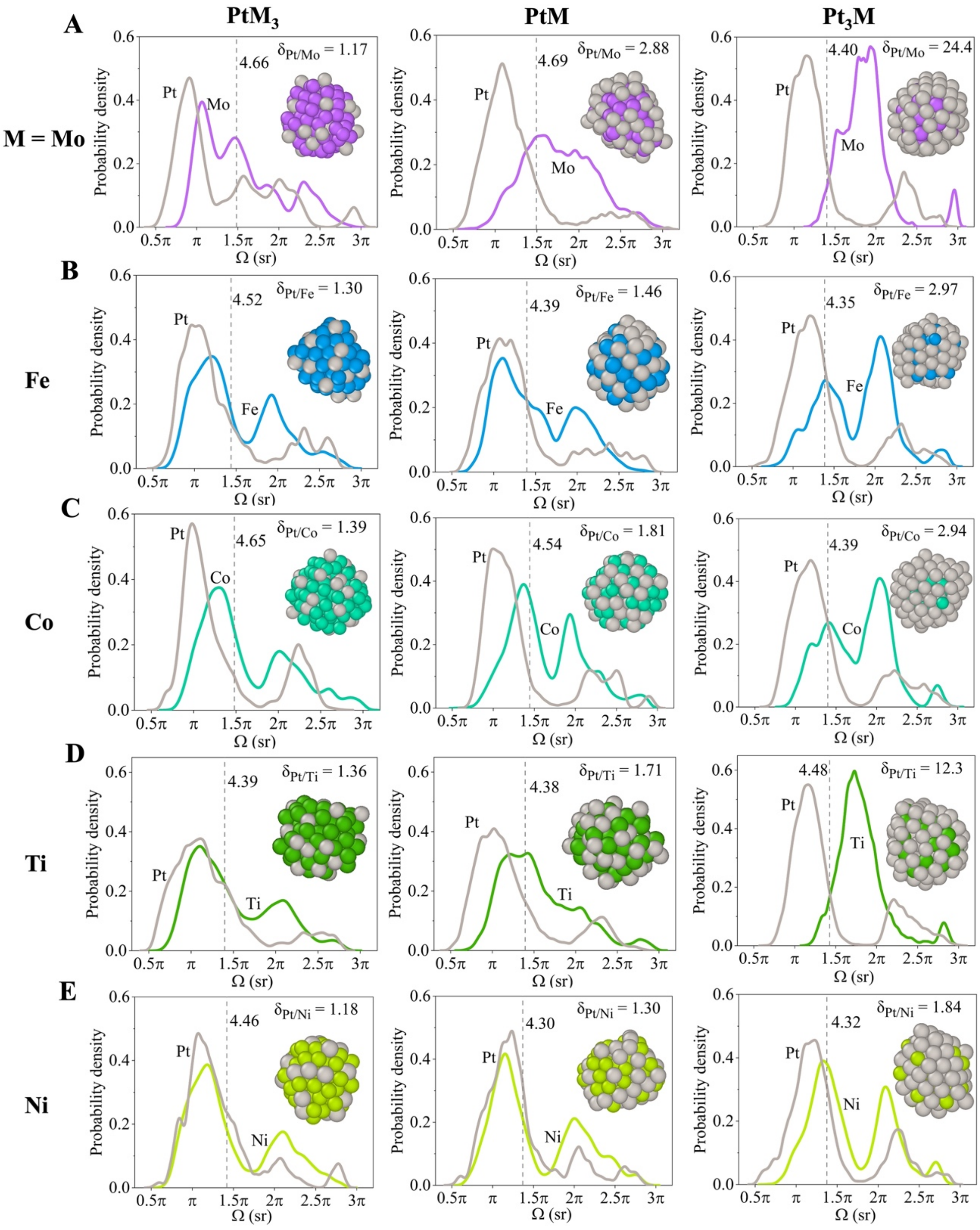

**Figure S6.** Structures at the end of MD simulations and solid angle distributions for 15 Pt-based bimetallic nanoparticles of different ratio corresponding to Figure 2. (A). PtMo system. (B). PtFe system. (C). PtCo system. (D). PtTi system. (E). PtNi system.

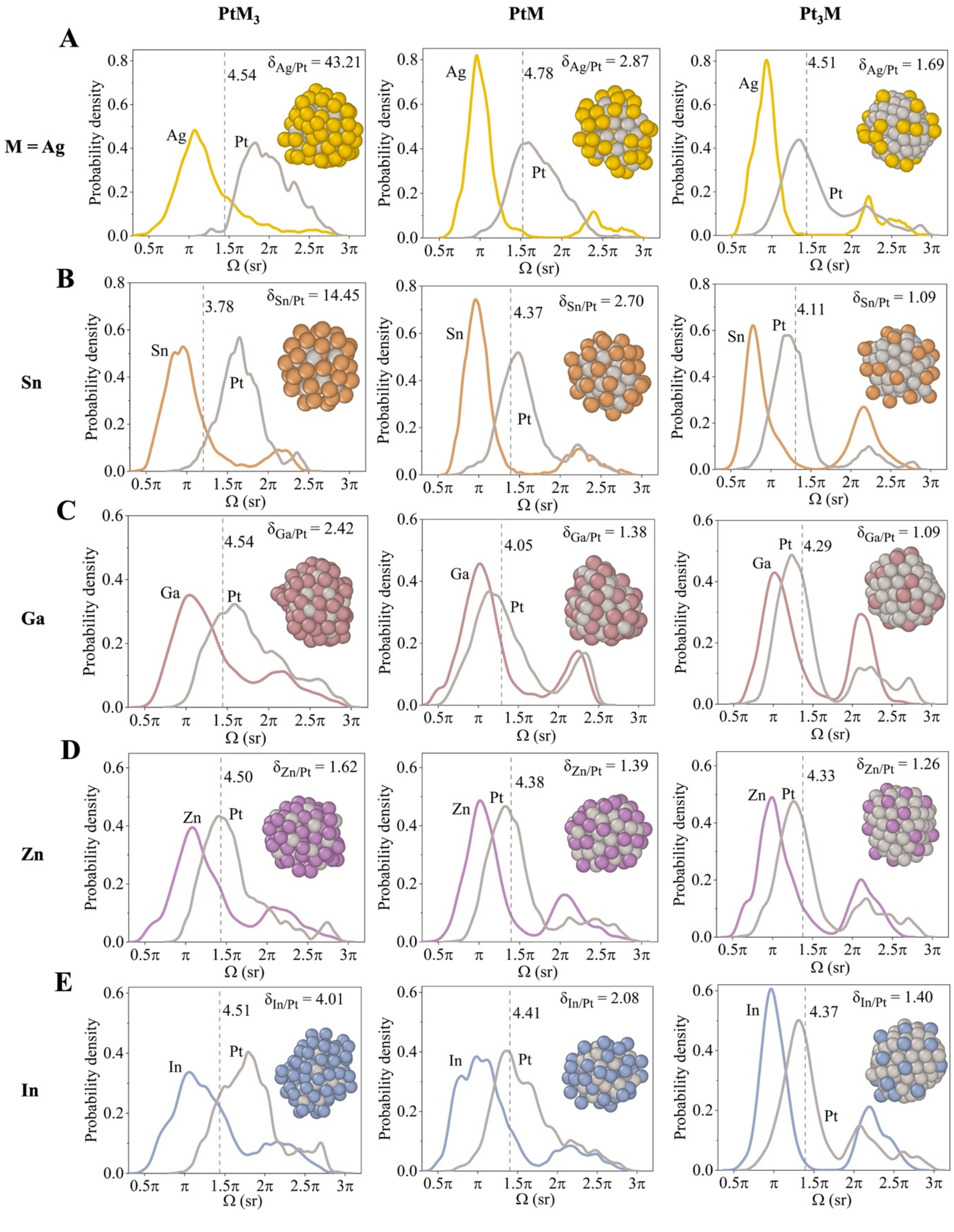

**Figure S7.** Structures at the end of MD simulations and solid angle distributions for 15 Pt-based bimetallic nanoparticles of different ratio corresponding to Figure 2. (A). PtAg system. (B). PtSn system. (C). PtGa system. (D). PtZn system. (E). PtIn system.

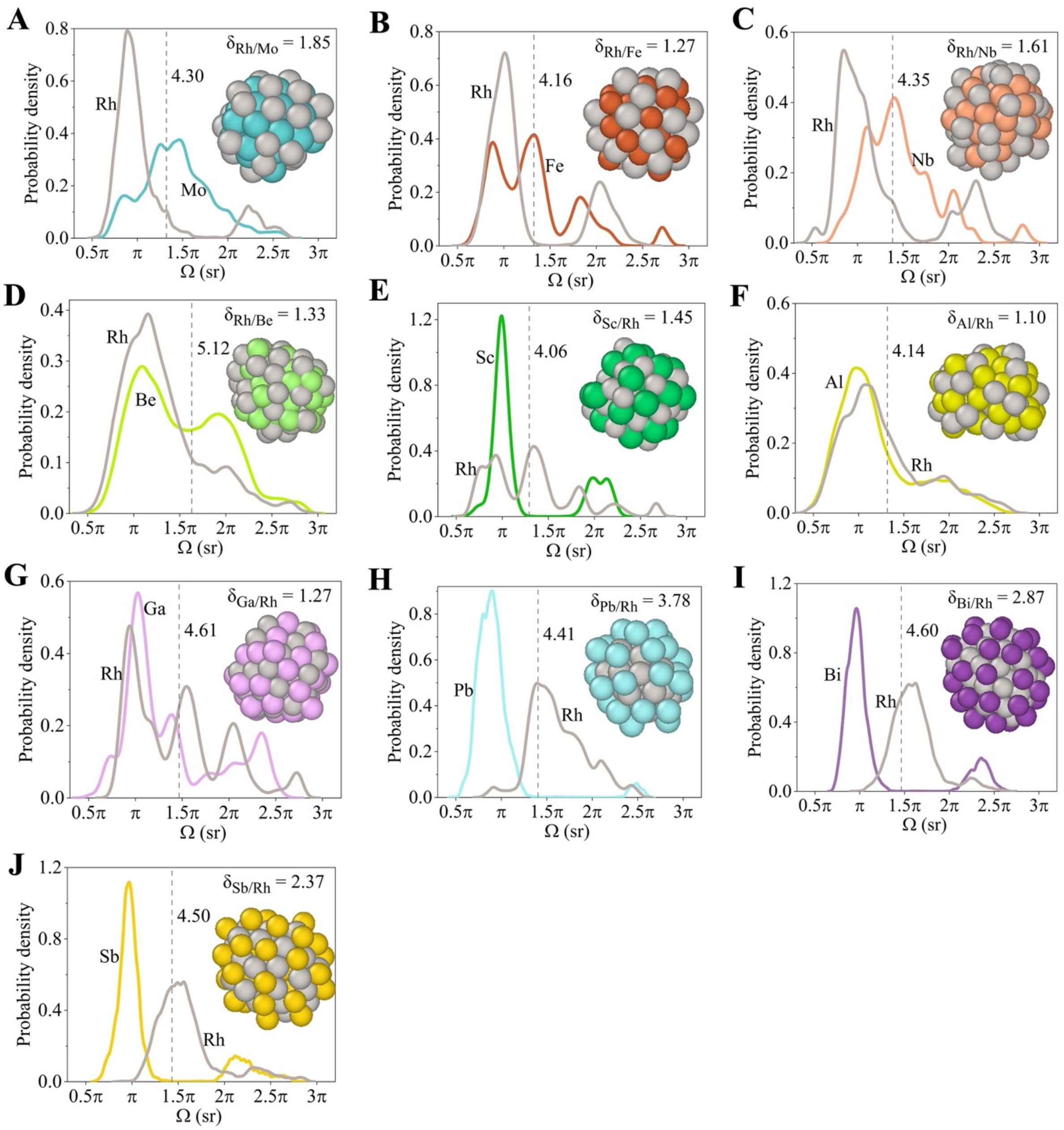

**Figure S8.** Structures at the end of MD simulations and solid angle distributions for 10 Rh-based bimetallic nanoparticles: (A) $Rh_{32}Mo_{32}$, (B) $Rh_{32}Fe_{32}$, (C) $Rh_{39}Nb_{39}$, (D) $Rh_{32}Be_{32}$, (E) $Rh_{32}Sc_{32}$, (F) $Rh_{32}Al_{32}$, (G) $Rh_{44}Ga_{44}$, (H) $Rh_{30}Pb_{30}$, (I) $Rh_{42}Bi_{42}$, and (J) $Rh_{40}Sb_{40}$.

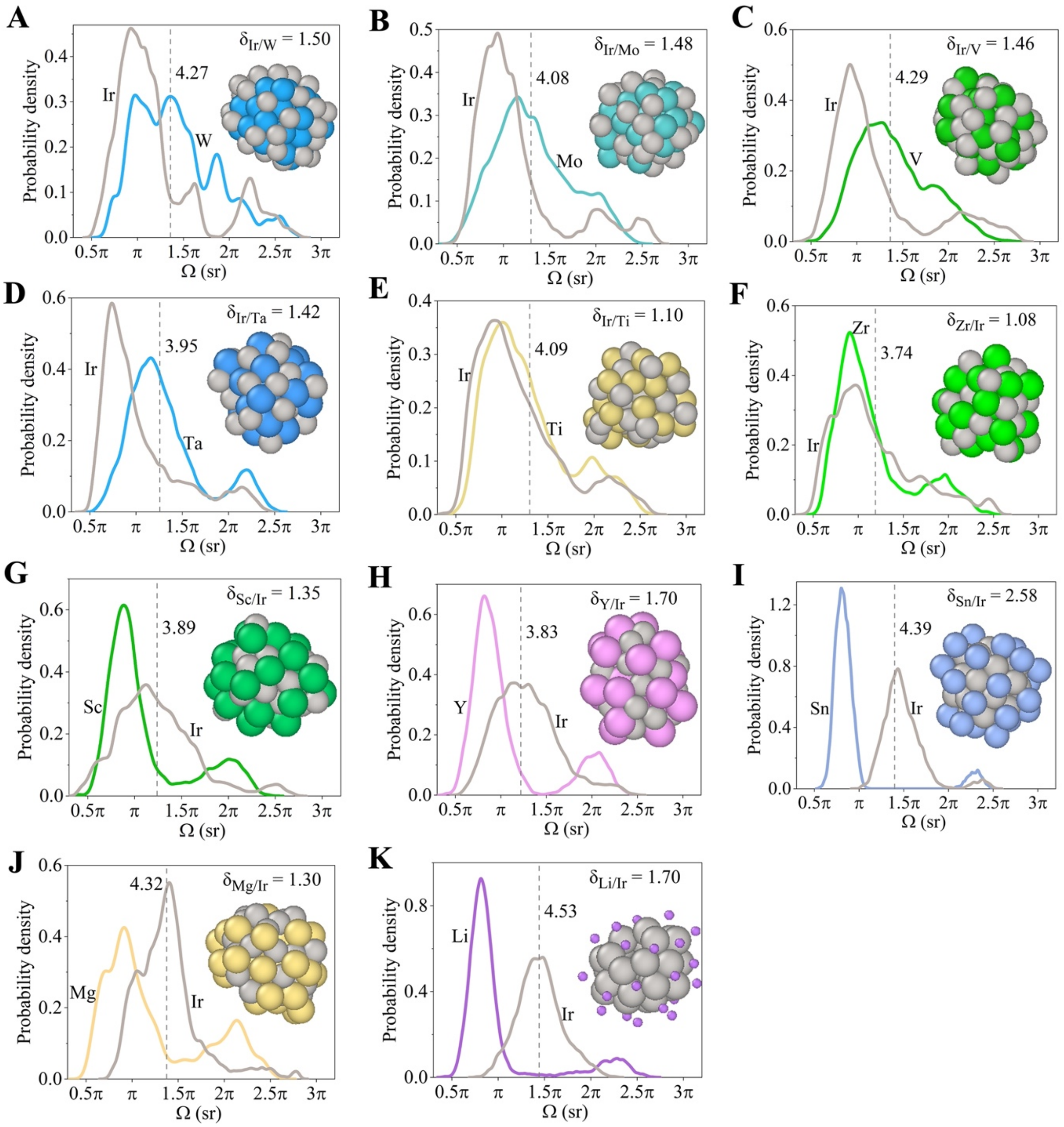

**Figure S9.** Structures at the end of MD simulations and solid angle distributions for 11 Ir-based bimetallic nanoparticles: (A) $Ir_{36}W_{36}$, (B) $Ir_{30}Mo_{30}$, (C) $Ir_{36}V_{36}$, (D) $Ir_{28}Ta_{28}$, (E) $Ir_{31}Ti_{31}$, (F) $Ir_{26}Zr_{26}$, (G) $Ir_{27}Sc_{27}$, (H) $Ir_{27}Y_{27}$, (I) $Ir_{26}Sn_{26}$, (J) $Ir_{36}Mg_{36}$, and (K) $Ir_{28}Li_{28}$.

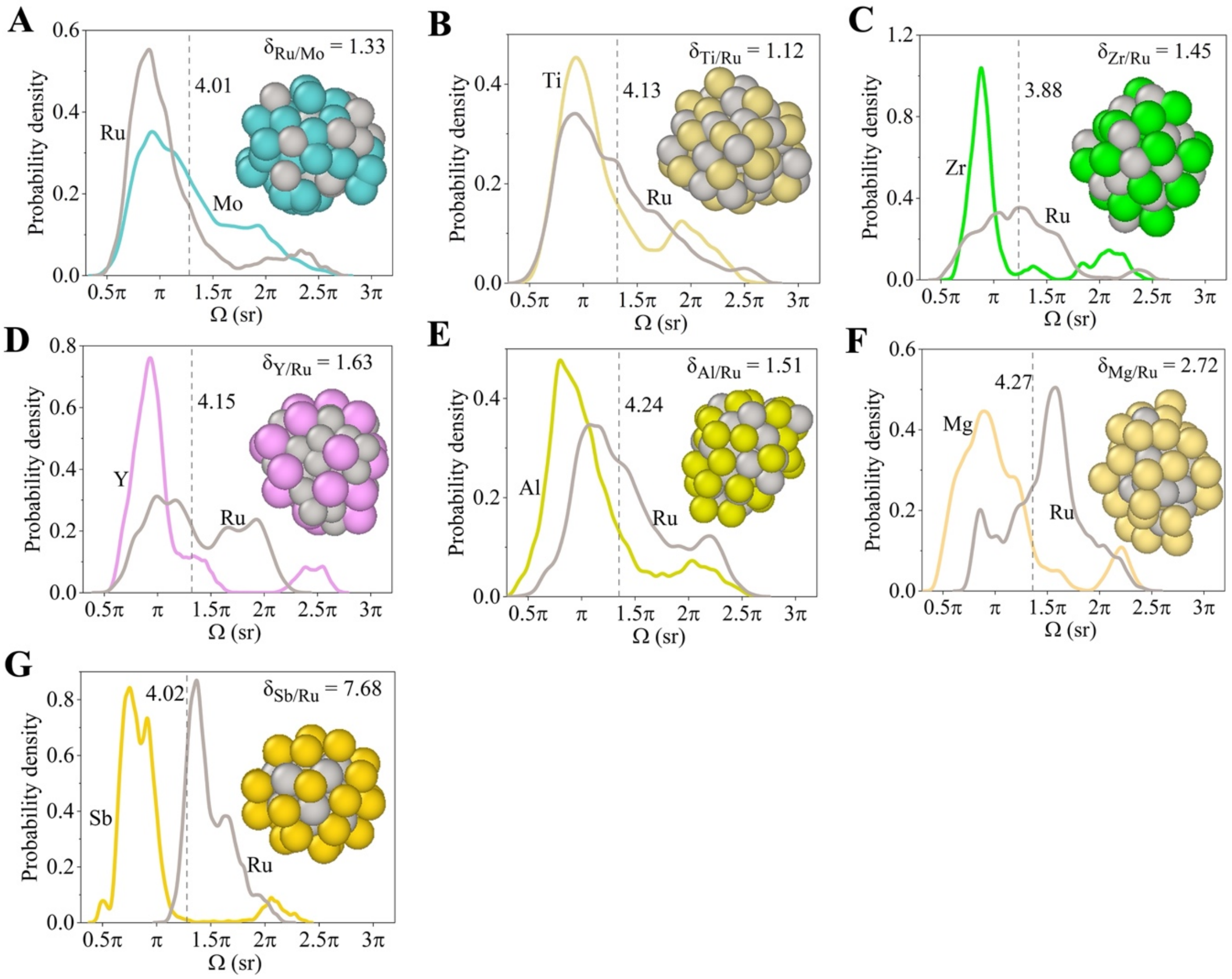

**Figure S10.** Structures at the end of MD simulations and solid angle distributions for 7 Ru-based bimetallic nanoparticles: (A) $Ru_{19}Mo_{38}$, (B) $Ru_{32}Ti_{32}$, (C) $Ru_{27}Zr_{27}$, (D) $Ru_{44}Y_{22}$, (E) $Ru_{32}Al_{32}$, (F) $Ru_{23}Mg_{36}$, and (G) $Ru_{14}Sb_{28}$.

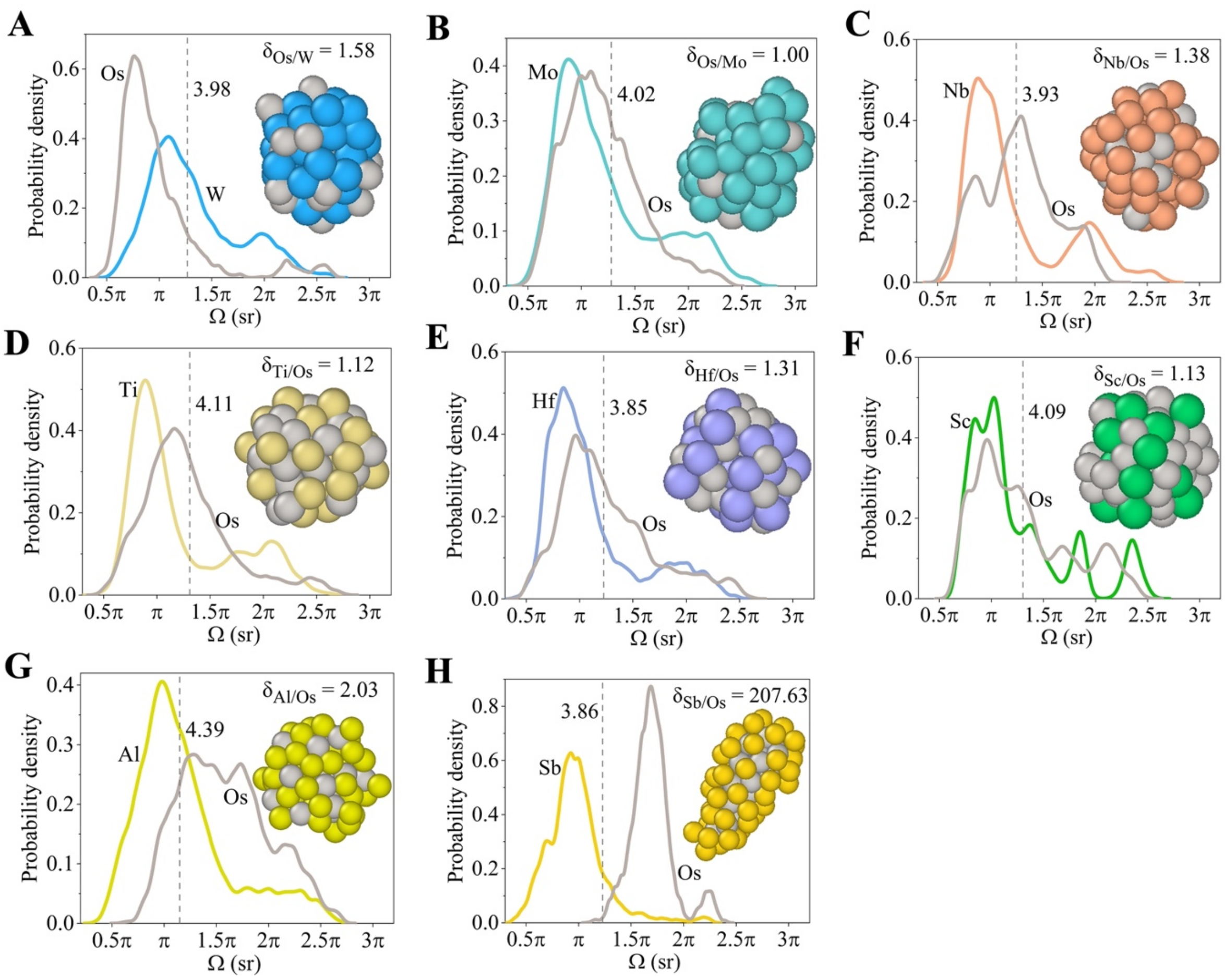

**Figure S11.** Structures at the end of MD simulations and solid angle distributions for 11 Os-based bimetallic nanoparticles: (A) $Os_{19}W_{38}$, (B) $Os_{15}Mo_{45}$, (C) $Os_{15}Nb_{45}$, (D) $Os_{32}Ti_{32}$, (E) $Os_{27}Hf_{27}$, (F) $Os_{44}Sc_{22}$, (G) $Os_{25}Al_{50}$, and (H) $Os_{23}Sb_{46}$.

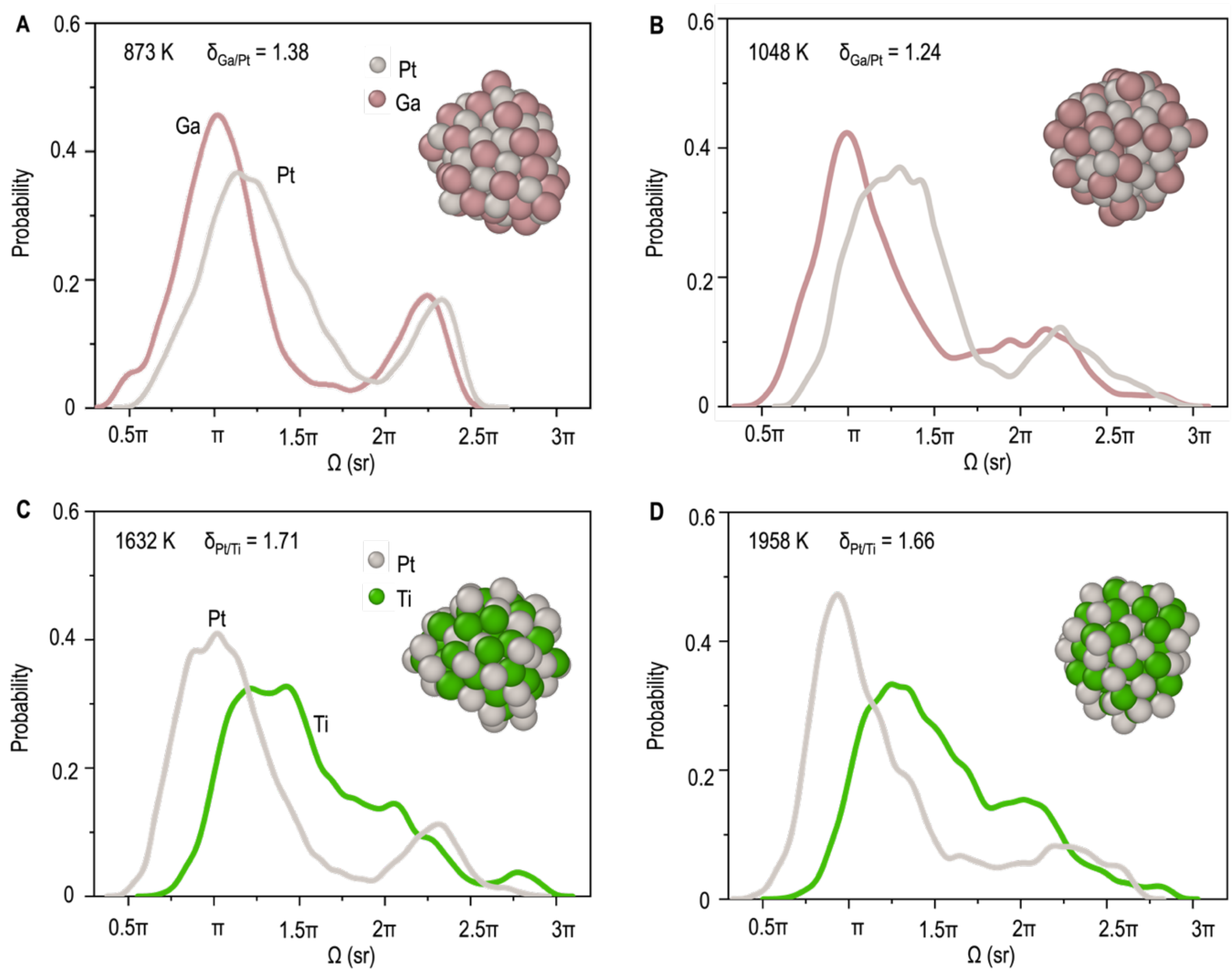

**Figure S12.** Temperature effects on surface segregation in the PtGa and PtTi nanoparticles. Equilibrated structures and solid angle distributions for PtGa at (A) 873 and (B) 1048 K, analyzed over 15,000 MD structures. Equilibrated structures and solid angle distributions for PtTi at (C) 1632 and (D) 1958 K, also analyzed over 15,000 MD structures.

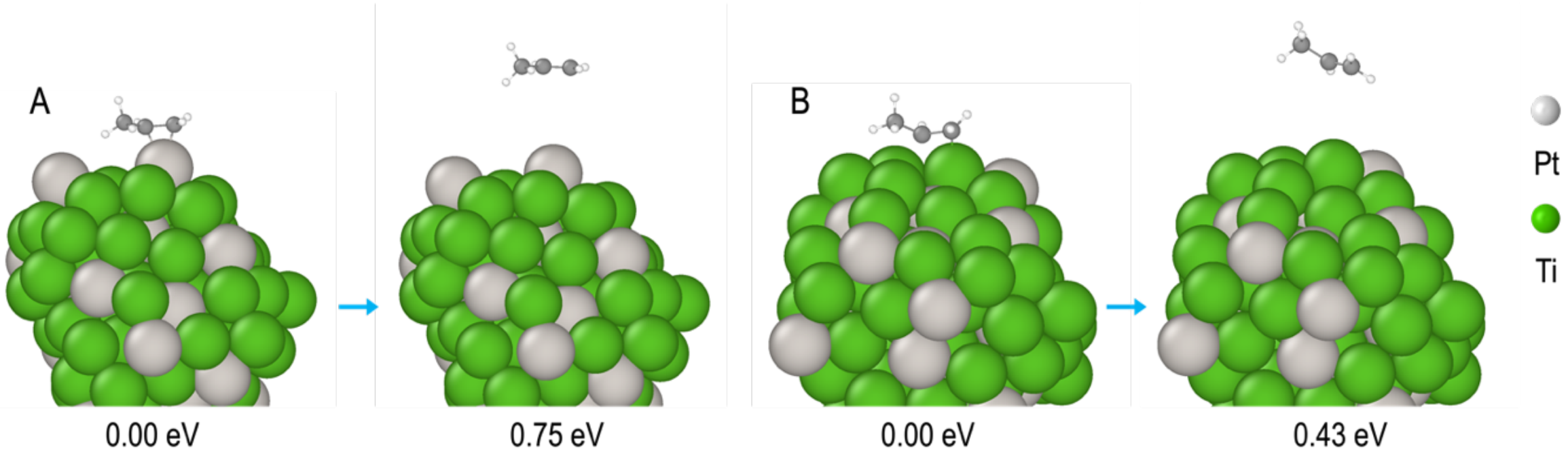

**Figure S13.** Structures and desorption energies of $C_3H_6^*$ on (A) Pt and (B) Ti sites in the $PtTi_3$ nanoparticle. The results show that the desorption energy on the Ti site is noticeably lower than that on the Pt site.

**Table S1.** Lowest surface energy of elemental bulk, $E_{surface}$, surface indices, formation energy of bimetallic alloy, $E_{alloy}$, and accumulation of element on nanoparticle surface. The surface energies and formation energy of bimetallic alloys are obtained from databases of The Materials Project[6] and Materials Virtual Lab.[15–17]

| *A* | $E^{A}_{surface}$ (J/m$^2$) | *B* | *Surface indices* | $E^{B}_{surface}$ (J/m$^2$) | Alloy | $E_{alloy}$ (eV/atom) | Surface Accumulation | $\delta$ |
|---|---|---|---|---|---|---|---|---|
| | | V | (100) | 2.38 | $Pd_2V$ | -0.27 | Pd | 5.72 |
| | | Ta | (110) | 2.34 | PdTa | -0.37 | Pd | 3.14 |
| | | Nb | (110) | 2.06 | $Pd_2Nb$ | -0.42 | Pd | 5.29 |
| | | Ti | $(11\bar{2}1)$ | 1.93 | $PdTi_2$ | -0.45 | Pd | 1.45 |
| | | Be | (0001) | 1.80 | PdBe | -0.59 | Pd | 1.49 |
| | | Hf | (0001) | 1.72 | $PdHf_2$ | -0.53 | Pd | 1.67 |
| | | Zr | $(10\bar{1}1)$ | 1.57 | PdZr | -0.69 | Pd | 1.36 |
| **Pd (111)** | 1.36 | Cu | (111) | 1.34 | PdCu | -0.12 | Pd | 1.09 |
| | | Sc | $(10\bar{1}0)$ | 1.20 | PdSc | -0.91 | NA | 1.02 |
| | | Al | (111) | 0.77 | PdAl | -0.94 | Pd | 1.06 |
| | | Ca | (0001) | 0.46 | PdCa | -0.65 | Ca | 2.03 |
| | | Zn | (0001) | 0.33 | PdZn | -0.58 | Zn | 1.27 |
| | | Pb | (111) | 0.26 | $PdPb_2$ | -0.21 | Pb | 27.01 |
| | | Bi | (0001) | 0.17 | PdBi | -0.35 | Bi | 2.96 |
| | | Cd | (0001) | 0.16 | PdCd | -0.39 | Cd | 1.97 |
| | | W | (110) | 3.23 | $Pt_2W$ | -0.35 | Pt | 7.02 |
| | | | | | $Pt_3Mo$ | -0.15 | Pt | 34.4 |
| | | Mo | (110) | 2.78 | PtMo | -0.36 | Pt | 2.95 |
| | | | | | $PtMo_3$ | -0.10 | Pt | 1.22 |
| | | | | | $Pt_3Fe$ | -0.17 | Pt | 3.12 |
| | | Fe | (110) | 2.45 | PtFe | -0.23 | Pt | 1.46 |
| | | | | | $PtFe_3$ | -0.07 | Pt | 1.34 |
| | | V | (100) | 2.38 | PtV | -0.55 | Pt | 1.53 |
| | | Ta | (110) | 2.34 | $Pt_2Ta$ | -0.74 | Pt | 3.69 |
| | | | | | $Pt_3Co$ | -0.05 | Pt | 3.04 |
| | | Co | (0001) | 2.11 | PtCo | -0.07 | Pt | 1.81 |
| | | | | | $PtCo_3$ | -0.08 | Pt | 1.39 |
| | | Nb | (110) | 2.06 | PtNb | -0.66 | Pt | 2.14 |
| | | | | | $Pt_3Ti$ | -0.92 | Pt | 12.34 |
| **Pt (111)** | 1.49 | Ti | $(11\bar{2}1)$ | 1.93 | PtTi | -0.87 | Pt | 1.71 |
| | | | | | $PtTi_3$ | -0.68 | Pt | 1.36 |
| | | | | | $Pt_3Ni$ | -0.06 | Pt | 1.94 |
| | | Ni | (111) | 1.92 | PtNi | -0.10 | Pt | 1.33 |
| | | | | | $PtNi_3$ | -0.07 | Pt | 1.23 |
| | | Hf | (0001) | 1.72 | PtHf | -1.26 | Pt | 1.40 |
| | | Zr | $(10\bar{1}1)$ | 1.57 | PtZr | -1.23 | Pt | 1.35 |
| | | Cu | (111) | 1.34 | PtCu | -0.18 | Cu | 1.11 |
| | | Al | (111) | 0.77 | PtAl | -1.13 | Pt | 1.05 |
| | | | | | $Pt_3Ag$ | 0.03 | Ag | 1.68 |
| | | Ag | (111) | 0.76 | PtAg | -0.08 | Ag | 2.90 |
| | | | | | $PtAg_3$ | -0.02 | Ag | 42.8 |
| | | | | | $Pt_3Sn$ | -0.40 | Sn | 1.09 |
| | | Sn | (221) | 0.54 | PtSn | -0.62 | Sn | 2.70 |
| | | | | | $PtSn_4$ | -0.23 | Sn | 14.45 |
| | | | | | $Pt_3Ga$ | -0.43 | Ga | 1.09 |
| | | Ga | (100) | 0.48 | PtGa | -0.63 | Ga | 1.38 |
| | | | | | $PtGa_3$ | -0.17 | Ga | 2.42 |
| | | | | | $Pt_3Zn$ | -0.31 | Zn | 1.23 |
| | | Zn | (0001) | 0.33 | PtZn | -0.57 | Zn | 1.39 |
| | | | | | $PtZn_3$ | -0.40 | Zn | 1.49 |
| | | Ba | (110) | 0.31 | PtBa | -0.75 | Ba | 1.96 |
| | | | | | $Pt_3In$ | -0.44 | In | 1.40 |
| | | In | (101) | 0.30 | PtIn | -0.57 | In | 2.05 |
| | | | | | $PtIn_3$ | -0.12 | In | 4.15 |
| | | Na | (0001) | 0.19 | $Pt_2Na$ | -0.46 | Na | 1.61 |
| | | Bi | (0001) | 0.17 | PtBi | -0.30 | Bi | 2.32 |
| | | Cd | (0001) | 0.16 | PtCd | -0.29 | Cd | 2.00 |
| | | Mo | (110) | 2.78 | RhMo | -0.19 | Rh | 1.85 |
| | | Fe | (110) | 2.45 | RhFe | -0.04 | Rh | 1.27 |
| | | Nb | (110) | 2.06 | RhNb | -0.40 | Rh | 1.61 |
| | | Be | (0001) | 1.80 | RhBe | -0.61 | Rh | 1.35 |
| **Rh (111)** | 1.98 | Sc | $(10\bar{1}0)$ | 1.20 | RhSc | -1.03 | Sc | 1.45 |
| | | Al | (111) | 0.77 | RhAl | -1.18 | Al | 1.10 |
| | | Ga | (100) | 0.48 | RhGa | -0.68 | Ga | 1.27 |
| | | Pb | (111) | 0.26 | RhPb | -0.14 | Pb | 3.78 |
| | | Bi | (0001) | 0.17 | RhBi | -0.19 | Bi | 2.87 |
| | | Sb | 0001 | 0.16 | RhSb | -0.53 | Sb | 2.37 |
| | | W | (110) | 3.23 | IrW | -0.30 | Ir | 1.50 |
| | | Mo | (110) | 2.78 | IrMo | -0.35 | Ir | 1.48 |
| | | V | (100) | 2.38 | IrV | -0.61 | Ir | 1.46 |
| | | Ta | (110) | 2.34 | IrTa | -0.65 | Ir | 1.42 |
| | | Ti | $(11\bar{2}1)$ | 1.93 | IrTi | -0.86 | Ir | 1.10 |
| **Ir (111)** | 2.36 | Zr | $(10\bar{1}1)$ | 1.57 | IrZr | -1.01 | Zr | 1.08 |
| | | Sc | $(10\bar{1}0)$ | 1.20 | IrSc | -1.06 | Sc | 1.35 |
| | | Y | $(10\bar{1}0)$ | 0.96 | IrY | -0.80 | Y | 1.70 |
| | | Sn | (221) | 0.54 | IrSn | -0.28 | Sn | 2.58 |
| | | Mg | (0001) | 0.51 | IrMg | -0.36 | Mg | 1.30 |
| | | Li | (100) | 0.46 | IrLi | -0.28 | Li | 1.70 |
| | | Mo | (110) | 2.78 | $RuMo_2$ | 0.02 | Ru | 1.33 |
| | | Ti | $(11\bar{2}1)$ | 1.93 | RuTi | -0.84 | Ti | 1.12 |
| | | Zr | $(10\bar{1}1)$ | 1.57 | RuZr | -0.74 | Zr | 1.45 |
| **Ru (0001)** | 2.60 | Y | $(10\bar{1}0)$ | 0.96 | $Ru_2Y$ | -0.30 | Y | 1.63 |
| | | Al | (111) | 0.77 | RuAl | -0.73 | Al | 1.51 |
| | | Mg | (0001) | 0.51 | $Ru_2Mg_3$ | -0.13 | Mg | 2.72 |
| | | Sb | (0001) | 0.16 | $RuSb_2$ | -0.29 | Sb | 7.68 |
| | | W | (110) | 3.23 | $OsW_2$ | 0.01 | Os | 1.58 |
| | | Mo | (110) | 2.78 | $OsMo_3$ | 0.00 | NA | 1.00 |
| | | Nb | (110) | 2.06 | $OsNb_3$ | -0.28 | Nb | 1.38 |
| **Os (0001)** | 2.95 | Ti | $(11\bar{2}1)$ | 1.93 | OsTi | -0.83 | Ti | 1.12 |
| | | Hf | (0001) | 1.72 | OsHf | -0.85 | Hf | 1.31 |
| | | Sc | $(10\bar{1}0)$ | 1.20 | $Os_2Sc$ | -0.38 | Os | 1.13 |
| | | Al | (111) | 0.77 | $OsAl_2$ | -0.67 | Al | 2.03 |
| | | Sb | (0001) | 0.16 | $OsSb_2$ | -0.08 | Sb | 207.63 |

**Table S2.** Charge state $q$ and $d$-band center $d_{center}$ of the examined surface Pt sites on the $PtTi_3$ nanoparticle discussed in Figure 4 of the main text.

| **Site** | ***q* (e)** | **$d_{center}$ (eV)** |
|---|---|---|
| **1** | -1.79 | -3.28 |
| **2** | -1.81 | -3.16 |
| **3** | -2.05 | -3.24 |
| **4** | -2.38 | -3.6 |
| **5** | -3.31 | -3.61 |
| **6** | -1.75 | -3.4 |
| **7** | -3.15 | -3.63 |
| **8** | -2.13 | -3.38 |
| **9** | -1.83 | -3.19 |
| **10** | -2.86 | -3.6 |
| **11** | -2.29 | -3.44 |
| **12** | -1.93 | -3.4 |

**Table S3.** Free Energies of intermediates and transition states along the reaction coordinate of propane dehydrogenation on different sites of $PtTi_3$ nanoparticle in Figure 4. All energies are referenced to isolated gas-phase reactants.

| | *+$C_3H_8$ | $C_3H_8$* | TS1 | $C_3H_7$*+H* | TS2 | $C_3H_6$*+2H* | $C_3H_6$*+$H_2$ | $C_3H_6$+$H_2$ |
|---|---|---|---|---|---|---|---|---|
| 1 | 0 | 1.04 | 1.79 | 0.33 | 1.24 | 0.42 | 0.81 | -0.06 |
| 2 | 0 | 0.82 | 2.07 | 0.78 | 1.51 | 0.11 | 0.79 | -0.06 |
| 3 | 0 | 0.99 | 2.1 | 0.8 | 1.28 | 0.2 | 0.89 | -0.06 |
| 4 | 0 | 1.27 | 2.22 | 0.53 | 1.56 | -0.02 | 0.67 | -0.06 |
| 5 | 0 | 1.35 | 2.27 | 0.68 | 1.64 | 0.42 | 1.1 | -0.06 |
| 6 | 0 | 1.2 | 2.29 | 0.62 | 1.78 | 0.15 | 0.83 | -0.06 |
| 7 | 0 | 1.17 | 2.31 | 0.48 | 1.06 | 0 | 0.68 | -0.06 |
| 8 | 0 | 0.98 | 2.01 | 0.34 | 1.43 | -0.43 | 0.25 | -0.06 |
| 9 | 0 | 1.36 | 2.52 | 0.81 | 1.57 | 0.35 | 1.04 | -0.06 |
| 10 | 0 | 1.4 | 2.52 | 0.45 | 1.73 | 0.22 | 0.91 | -0.06 |
| 11 | 0 | 1.32 | 2.44 | 0.56 | 1.57 | -0.26 | 0.42 | -0.06 |
| 12 | 0 | 1.2 | 2.55 | 0.78 | 1.88 | 0.27 | 0.9 | -0.06 |

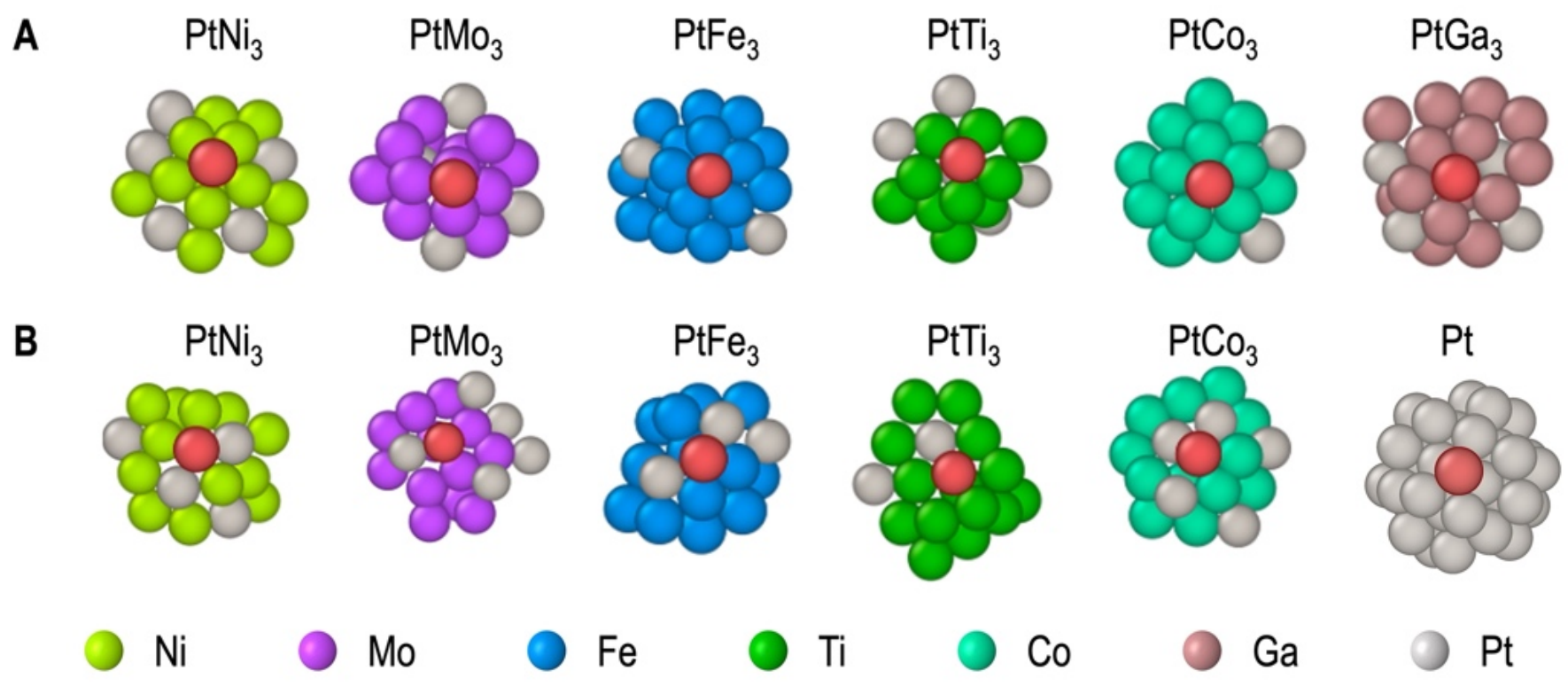

**Figure S14.** Configurations of the examined (A) isolated Pt sites and (B) Pt ensemble sites in Figure 5 of the main text. The examined Pt sites are highlighted in red.

**Table S4.** Free Energies of intermediates and transition states along the reaction coordinate of propane dehydrogenation on different nanoparticles examined in Figure 5. All energies are referenced to the initial reactants.

| | | **$*+C_3H_8$** | **$C_3H_8*$** | **TS1** | **$C_3H_7*+H*$** | **TS2** | **$C_3H_6*+2H*$** | **$C_3H_6*+H_2$** | **$C_3H_6+H_2$** |
|---|---|---|---|---|---|---|---|---|---|
| Pt ensemble site | $PtNi_3$ | 0 | 1.12 | 1.65 | 0.74 | 1.35 | 0.61 | 0.23 | -0.06 |
| | $PtTi_3$ | 0 | 1.32 | 2.44 | 0.56 | 1.57 | -0.26 | 0.42 | -0.06 |
| | $PtCo_3$ | 0 | 1.14 | 1.77 | 0.97 | 1.58 | 0.77 | 0.4 | -0.06 |
| | $PtFe_3$ | 0 | 1.3 | 2.36 | 0.98 | 1.3 | 0.81 | 0.4 | -0.06 |
| | $PtMo_3$ | 0 | 1.05 | 1.65 | 0.9 | 1.46 | 0.38 | -0.14 | -0.06 |
| | Pt | 0 | 0.97 | 1.47 | 0.31 | 0.97 | -0.73 | -0.29 | -0.06 |
| Pt isolated site | $PtNi_3$ | 0 | 1.03 | 1.66 | 1 | 1.31 | 0.63 | 0.47 | -0.06 |
| | $PtTi_3$ | 0 | 1.04 | 1.79 | 0.33 | 1.24 | 0.42 | 0.81 | -0.06 |
| | $PtCo_3$ | 0 | 1.38 | 1.97 | 1.07 | 1.62 | 0.87 | 0.54 | -0.06 |
| | $PtFe_3$ | 0 | 1.23 | 1.8 | 1.06 | 2 | 0.7 | 0.35 | -0.06 |
| | $PtMo_3$ | 0 | 1.15 | 2.64 | 1.82 | 2.35 | 1.67 | 1.02 | -0.06 |
| | $PtGa_3$ | 0 | 1.3 | 2.76 | 1.9 | 3.32 | 2.06 | 0.87 | -0.06 |

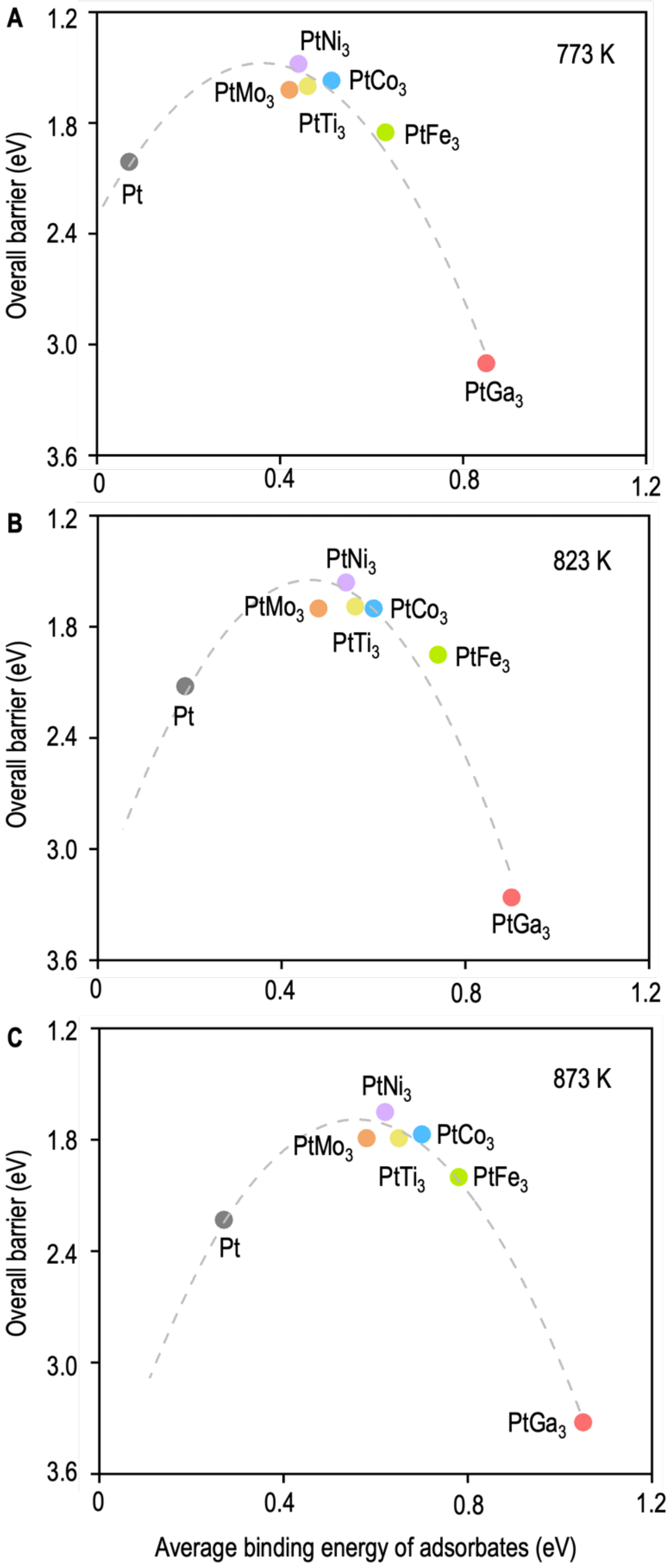

**Figure S15.** Correlation between overall energy barriers and average binding energy of adsorbates ($C_3H_8$, $C_3H_7$, $C_3H_6$, and H) at (A) 773, (B) 823, and (C) 873 K.